\documentclass[a4paper,10pt]{article}

\usepackage[pdftex]{color,graphicx,hyperref}
\usepackage{amsmath,amsfonts,amssymb}
\usepackage[margin=1in]{geometry}
\usepackage{setspace}
\usepackage{booktabs}
\usepackage[labelfont=bf,labelsep=period]{caption}
\usepackage{subcaption}
\usepackage{cite}
\usepackage[utf8]{inputenc}
\usepackage[affil-it]{authblk}

\newcommand{\eref}[1]{Eq.~(\ref{#1})}

\newcommand{\fref}[1]{Fig.~\ref{#1}}

\newcommand{\srefsi}[1]{SI Section~S{#1}}
\newcommand{\frefsi}[1]{SI Fig.~S{#1}}

\newcommand{\trefsi}[1]{SI Table~S{#1}}
\newcommand{\tsrefsi}[2]{SI Tables~S{#1}-S{#2}}

\providecommand{\keywords}[1]{{\small {\bf Keywords---} #1}}

\title{Universal patterns in egocentric communication networks}

\author[1,2,3,4,*]{Gerardo Iñiguez}
\author[2]{Sara Heydari}
\author[1,5]{János Kertész}
\author[2*]{Jari Saramäki}

\affil[1]{\small{Department of Network and Data Science, Central European University, 1100 Vienna, Austria}}
\affil[2]{\small{Department of Computer Science, Aalto University School of Science, 00076 Aalto, Finland}}
\affil[3]{\small{Faculty of Information Technology and Communication Sciences, Tampere University, 33720 Tampere, Finland}}
\affil[4]{\small{Centro de Ciencias de la Complejidad, Universidad Nacional Auton\'{o}ma de M\'{e}xico, 04510 Ciudad de México, Mexico}}
\affil[5]{\small{Complexity Science Hub, 1080 Vienna, Austria}}
\affil[*]{\small{Corresponding author email: iniguezg@ceu.edu, jari.saramaki@aalto.fi}}

\date{}

\begin{document}

\maketitle

\begin{abstract}
Tie strengths in social networks are heterogeneous, with strong and weak ties playing different roles at both the network and the individual level. Egocentric networks, networks of relationships around a focal individual, exhibit a small number of strong ties and a larger number of weaker ties, a pattern that is evident in electronic communication records, such as mobile phone calls. Mobile phone data has also revealed persistent individual differences within this pattern. However, the generality and the driving mechanisms of this tie strength heterogeneity remain unclear. Here, we study tie strengths in egocentric networks across multiple datasets containing records of interactions between millions of people over time periods ranging from months to years. Our findings reveal a remarkable universality in the distribution of tie strengths and their individual-level variation across different modes of communication, even in channels that may not reflect offline social relationships. With the help of an analytically tractable model of egocentric network evolution, we show that the observed universality can be attributed to the competition between cumulative advantage and random choice, two general mechanisms of tie reinforcement whose balance determines the amount of heterogeneity in tie strengths. Our results provide new insights into the driving mechanisms of tie strength heterogeneity in social networks and have implications for the understanding of social network structure and individual behavior.
\end{abstract}

\keywords{Quantitative sociology, personal networks, tie strength}

\section*{Introduction}

Social networks are key to the exchange of ideas, norms, and other cultural constructs in human society~\cite{wasserman1994social}, influencing the way we communicate~\cite{tomasello2010origins}, support each other~\cite{house1988social,holt2010social}, and form enduring communities~\cite{wellman1979community}. Decades of research have focused on regularities in the patterns of relations among individuals~\cite{nadel1957theory} as well as the drivers and mechanisms behind their origin~\cite{borgatti2009network}. One particularly prominent feature of social networks is the diversity of tie strengths~\cite{granovetter1973strength}, where strong ties are typically embedded within social groups while weak ties are crucial for the cohesiveness of the network as a whole~\cite{granovetter1973strength,onnela2007structure,csermely2006weak}. At the micro level, ego networks---the sets of social ties between an individual (the ego) and their family, friends, and acquaintances (the alters)---commonly feature a small core of close relationships. These close relationships are associated with high emotional intensity and they are surrounded by a larger number of weaker ties. The emergence of this characteristic structural pattern has been associated with constraints on maintaining social relationships, which include limited information processing capacity~\cite{miller1956magical}, social cognition~\cite{bernard1973social,dunbar1998social,tamarit2018cognitive}, and time availability~\cite{gonccalves2011modeling,miritello2013time,lerman2016information}.

Studies of human communication via mobile phones have shown that in line with the above picture, there is a consistent, general pattern in  egocentric networks where a small number of close alters receive a disproportionately large share of communication. Data on the frequency of mobile phone calls and text messages also indicate that within this general pattern, there are clear and persistent individual differences~\cite{saramaki2014persistence,aledavood2016channel,centellegher2017personality,heydari2018multichannel,LiSignatures}: some people repeatedly focus most of their attention on a few close relationships, while others tend to distribute communication among their alters more evenly~\cite{saramaki2014persistence}. These differences are stable in time even under high personal network turnover. However, the mechanisms that generate such heterogeneity of tie strengths, its individual-level variation, and the generality of this pattern beyond mobile-phone-mediated communication, have not yet been established~\cite{tamarit2018cognitive,LI2018213,LiSignatures,KOLTSOVA2021106856}. 

Here, we explore multiple sets of data on recurring social interactions between millions of people to study heterogeneity in ego network tie strengths and its individual variation, and to shed light on the mechanisms behind this heterogeneity. These large-scale data sets contain metadata on different types of time-stamped interactions, from mobile phone calls to social media, spanning a time range from months to years. They are likely to reflect different aspects of social behaviour: e.g., mobile-phone calls between friends, work-related emails, and messages on an Internet forum or dating website serve different purposes and may or may not reflect social relationships that also exist offline.
Using social networks reconstructed from the interaction records in our data, we measure the distribution of tie strengths in a massive number of egocentric networks, focusing on how this distribution varies between individuals. We compare observations across several datasets representing different channels of communication and use our observations to construct a minimal, analytically tractable model of egocentric network growth that attributes heterogeneity in tie strengths and its individual variation to the balance between competing mechanisms of tie reinforcement. 

We find systematic evidence of broad variation in the distributions of tie strengths in ego networks across all communication channels, including those channels that do not necessarily reflect offline social interactions. The majority of ego networks have heterogeneous tie strengths with varying amounts of heterogeneity, while a minority of individuals distribute their contacts in a homogeneous way. With the help of our model of egocentric network evolution, we attribute the amount of heterogeneity to a mechanism of cumulative advantage~\cite{merton1968matthew,price1976general,diprete2006cumulative}, similar to proportional growth~\cite{simon1955class} and preferential attachment~\cite{barabasi1999emergence,barrat2004weighted,toivonen2009comparative,karimi2018homophily}. Homogeneity, in turn, is associated with effectively random choice of alters for communication. The balance between these two mechanisms determines the dispersion of tie strengths in an egocentric network. This balance is captured in our model through a single preferentiality parameter that can be fitted to data for each ego. The distributions of fitted values of this parameter are remarkably similar across different datasets, indicating universal patterns of communication in channels that are very different in nature. Similarly to social signatures~\cite{saramaki2014persistence}, we also observe that at the level of individuals, the preferentiality parameter is a stable and persistent indicator of the distinctive way people shape their network on the particular channel.

\section*{Results} 

\begin{figure}[t]
\centering
\includegraphics[width=0.7\textwidth]{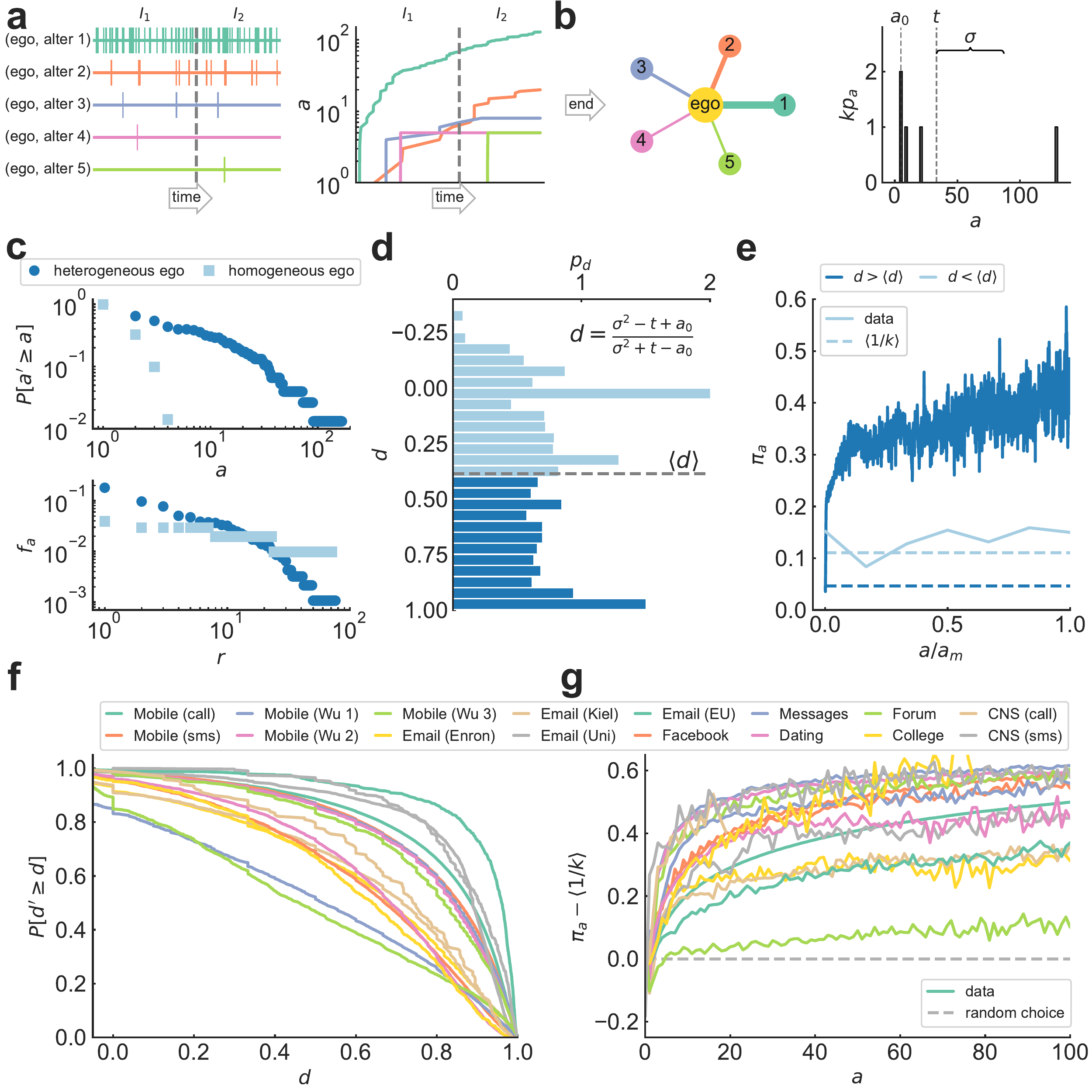}
\caption{\scriptsize {\bf Tie strengths in egocentric network are heterogeneous and driven by cumulative advantage.}
{\bf (a)} Real-time contact sequence between an ego and its $k$ alters (left) and time evolution of communication activity $a$ with each alter (right),  for selected ego in the CNS call dataset~\cite{stopczynski2014measuring,sapiezynski2019interaction} (for data description see \srefsi{1}).  Times are relative to the length of the observation period, so close-by events appear as single lines (left) or sudden increases in $a$ (right).  The sequence is divided into two consecutive intervals with the same number of events ($I_1$ and $I_2$). As time goes by, some alters accumulate more events than others.  {\bf (b)} Aggregated ego network (left) and final alter communication activity distribution $p_a$ (right) for data in (a). The distribution is characterized by a minimum activity $a_0$, mean $t$, and standard deviation $\sigma$.  {\bf (c)} Complementary cumulative distribution $P[a' \geq a]$ of number of alters with at least activity $a$ (top) and fraction of communication events $f_a$ with alter at rank $r$ (bottom), for selected egos in the Forum dataset~\cite{said2010social,karimi2014structural,saramaki2015exploring}.  Egos distribute activity among alters either homogeneously or heterogeneously.  {\bf (d)} Distribution $p_d$ of the dispersion index $d$ for all egos with more than 5 events in the Forum dataset, showing the systematic presence of both types of egos in (c).  {\bf (e)} Probability $\pi_a$ that an alter with activity $a$ is contacted, averaged over time and subsets of heterogeneous ($d > \langle d \rangle$) or homogeneous ($d < \langle d \rangle$) egos in (d),  and the average baseline $\pi_a = \langle 1/k \rangle$ when communication events are distributed randomly (each value of $a$ corresponds to at least 50 egos and is normalized by the maximum activity $a_m$ in the subset).  For heterogeneous ego networks, the increasing tendency indicates cumulative advantage where alters with high prior activity receive more communication.  {\bf (f)} Complementary cumulative distribution $P[d' \geq d]$ of the number of egos having at least dispersion $d$,  for 8.6M egos in 16 datasets of calls,  messaging,  and online interactions (\trefsi{1} and \frefsi{2}; shown only for ego networks with more than 10 events).  Data shows a broad variation in how egos allocate activity among alters.  {\bf (g)} Relative connection kernel $\pi_a - \langle 1/k \rangle$ for all datasets (each $a$ value corresponds to at least 50 egos with $k \geq 2$; see \frefsi{3}).  Increasing trends indicate cumulative advantage in the ego networks of all channels.
}
\label{fig:figure1}
\end{figure}

We analyze data on recurring, time-stamped social interactions between millions of individuals across 16 communication channels, including phone call records, text messages, emails, and posts from social networks and online forums (\fref{fig:figure1}). Data include, among others, anonymized metadata for 1.3B calls and 613M messages made by 6M people in a European country during 2007~\cite{onnela2007analysis,onnela2007structure,karsai2011small,kivela2012multiscale,kovanen2013temporal,unicomb2018threshold,heydari2018multichannel},  431k emails by 57k students at Kiel University in 4 months~\cite{ebel2002scale,saramaki2015exploring}, and 850k wall posts in Facebook made by 45k users in New Orleans during 2006--2009~\cite{viswanath2009evolution,saramaki2015exploring}.  Periods of observation vary widely,  from 1 month of text message logs for 3 mobile phone companies~\cite{wu2010evidence} to 7 years of private messages and open forum discussions in the Swedish movie recommendation website Filmtipset~\cite{said2010social,karimi2014structural,saramaki2015exploring} (for data details see Supplementary Information [SI] Section S1, Table S1, and Fig. S1).  The analyzed data covers a wide range of population sizes and time scales of activity, and they come from a large enough variety of channels to include typical social contexts of human online communication.

The total communication activity $a$ (the number of calls, messages, or posts) between an individual, or ego, and each of the ego's acquaintances,  or alters,  increases with time (\fref{fig:figure1}a). Due to variability in the communication patterns with different alters, aggregated ego networks at the end of the observation period typically have heterogeneous tie strengths (numbers of events between ego and alter), manifested as a broad alter activity distribution $p_a$.  Following~\cite{goh2008burstiness}, we characterize the spread of $p_a$ by the dispersion index $d = (\sigma^2 - t_r) / (\sigma^2 + t_r)$, where $\sigma^2$ is the variance of $p_a$ and $t_r = t - a_0$ its mean relative to the minimum activity in the ego network (\fref{fig:figure1}b).  We find that in our datasets most egos primarily communicate with a few alters,  in agreement with previously observed patterns of mobile phone communication~\cite{reid2005textmates,saramaki2014persistence} and online platform use~\cite{arnaboldi2013egocentric}. These egos have networks with heterogeneous tie strengths, in other words, broad activity distributions $p_a$ with large dispersion $d$, or equivalently, steep activity-rank curves (``social signatures" in~\cite{saramaki2014persistence}) where most events are concentrated on the highest-ranking alters~\cite{saramaki2014persistence,iniguez2022dynamics} (\fref{fig:figure1}c). Note that in the following, because of the equivalence, we use the term social signature interchangeably for both individual activity distributions and activity-rank curves.  In addition to egos with heterogeneous tie strengths, all studied communication channels contain a smaller fraction of egos who distribute their communication more homogeneously among alters, leading to smaller values of $d$ and flatter activity-rank distributions.  Indeed, the distribution $p_d$ of the dispersion indices over an entire dataset shows both over-dispersed egos ($d \sim 1$) and egos with more Poissonian social signatures ($d \sim 0$; \fref{fig:figure1}d).  Even egos with similar degrees or strength (total numbers of alters or events) can have heterogeneous or homogeneous activity distributions, which are thus not solely driven by differences in the total level of activity between individuals.

In order to find plausible generative mechanisms behind the diversity of social signatures seen in human communication data, we calculate the probability $\pi_a$ that a new contact happens between the ego and an alter with activity $a$, averaged over all events and alters in the aggregated ego network (\fref{fig:figure1}e).  This measure is akin to the attachment kernel of growing networks~\cite{newman2001clustering,jeong2003measuring,pham2015pafit}, which has been identified in many cases as a linear function of the degree \cite{krapivsky2001organization,eriksen2001scale}, and which has been applied in preferential attachment models~\cite{simon1955class,barabasi1999emergence,barrat2004weighted,bianconi2001competition}.  When averaged over heterogeneous egos ($d > \langle d \rangle$), $\pi_a$ increases roughly linearly with $a$, indicating cumulative advantage or linear growth as the way most individuals interact with their acquaintances. Homogeneous egos ($d < \langle d \rangle$), on the other hand, are closer to the average baseline $\pi_a = \langle 1/k \rangle$ where events are allocated among alters uniformly, which can be modelled by random choice. Despite variations in the ratio of heterogeneous to homogeneous activity distributions across channels (signaled by different shapes of the dispersion distribution $p_d$; \fref{fig:figure1}f and \frefsi{2}), the connection probability $\pi_a$ has qualitatively the same functional form for all datasets,  and it even has a similar slope for a wide range of activity values (\fref{fig:figure1}g and \frefsi{3}). 

\begin{figure}[t]
\centering
\includegraphics[width=0.8\textwidth]{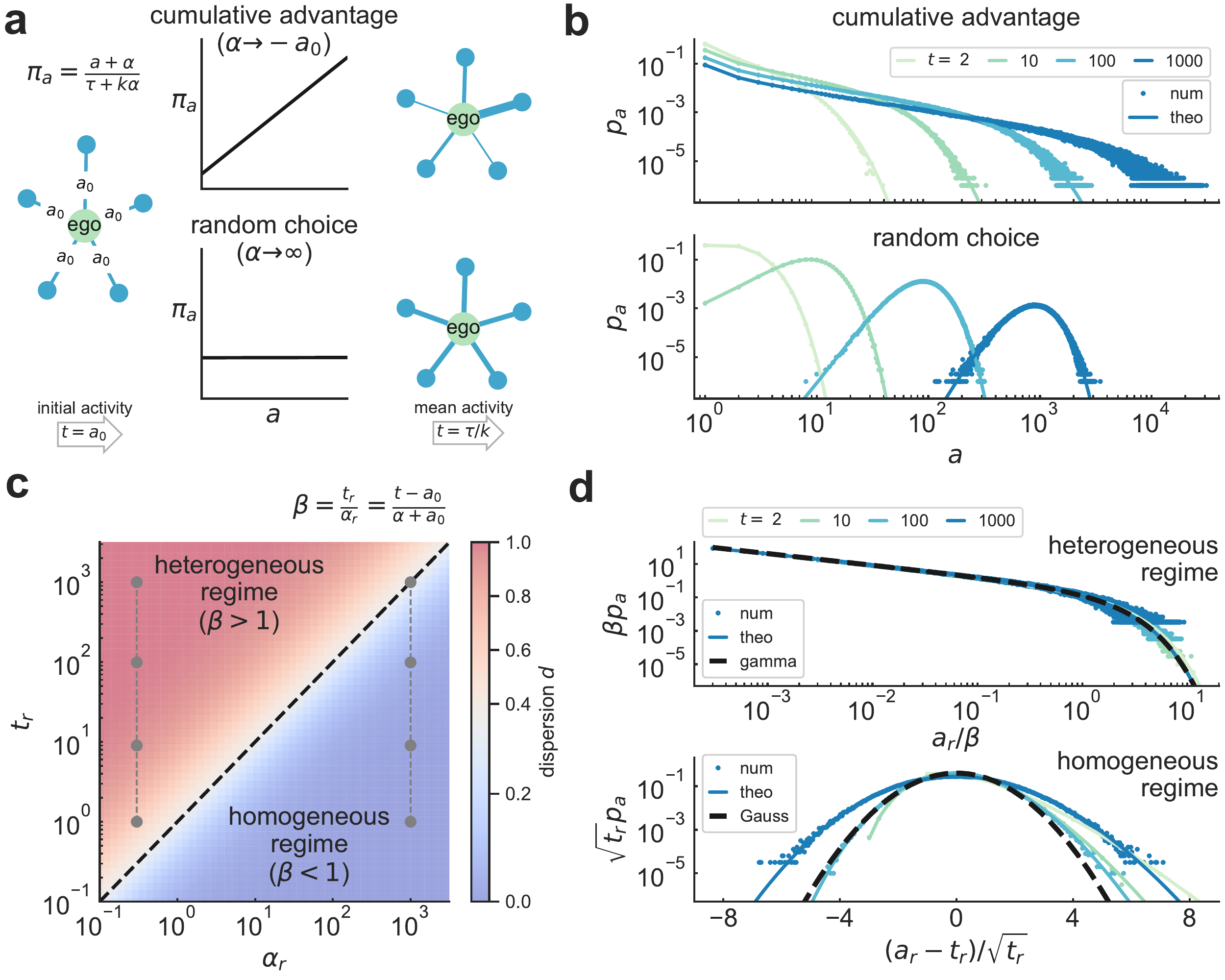}
\caption{\scriptsize {\bf Simple model of alter activity shows crossover in shape of social signatures.}
{\bf (a)} In a modeled ego network of degree $k$,  alters begin with activity $a_0$ and engage in new communication events at event time $\tau$ with probability $\pi_a$, where $a$ is the alter's current activity and $\alpha$ a parameter interpolating behavior between cumulative advantage ($\alpha \to -a_0$, top) and random choice ($\alpha \to \infty$, bottom; see MM and \srefsi{2}). These dynamics lead to an ego network with mean alter activity (i.e. time) $t = \tau / k$.  Plots and networks on the right are shown diagrammatically but correspond to $k=5$,  $a_0=1$, $\alpha=-0.9$ ($10^3$),  and $t=3$ ($10^3$) at the top (bottom). {\bf (b)} Probability $p_a$ that an alter has activity $a$ at time $t$, for varying $t$ with $\alpha = -0.7$ ($9$) at the top (bottom),  $k=100$ and $a_0=1$.  Numerical simulations (num) match well with analytical calculations (theo), indicating that cumulative advantage and random choice, respectively, lead to broad or narrow activity distributions.  {\bf (c)} Phase diagram of activity dispersion $d$ in terms of rescaled parameters $\alpha_r = \alpha + a_0$ and $t_r = t - a_0$.  The preferentiality parameter $\beta = t_r / \alpha_r$ showcases a crossover between heterogeneous and homogeneous regimes at $\beta=1$ (dashed line).  The vertical gray dash-dotted lines are parameter values for plot (d).  {\bf (d)} Rescaled activity distribution $p_a$ for varying $t$ and $\alpha_r = 0.3$ ($10^3$) at the top (bottom). Heterogeneous (homogeneous) regimes show gamma (Gaussian) scaling in $p_a$.  All simulations are averages over $10^4$ realizations.
}
\label{fig:figure2}
\end{figure}

To explore the simplest theoretical mechanisms that may give rise to the observed variability across ego networks, we consider minimal cumulative-advantage dynamics similar to Price's model~\cite{price1965networks,price1976general}, where the probability of communication between an ego and an alter depends on their prior communication activity and a tunable parameter $\alpha$ that modulates random alter choice (\fref{fig:figure2}).  We start with an undirected ego network of degree $k$ where all alters have initial communication activity $a_0$. After $\tau$ interactions,  the probability $\pi_a$ that an alter with activity $a$ interacts with the ego at event time $\tau+1$ is
\begin{equation}
\label{eq:connProb}
\pi_a = \frac{a + \alpha}{ \tau + k \alpha }.
\end{equation}
When the parameter $\alpha$ is small,  $\pi_a$ increases linearly with activity so egos interact preferentially with the most active alters, following a dynamics similar to stochastic processes driven by cumulative advantage~\cite{simon1955class,diprete2006cumulative}, and preferential attachment in the evolution of connectivity~\cite{barabasi1999emergence,bianconi2001competition,karimi2018homophily} and edge weights~\cite{barrat2004weighted} in growing networks.  For large $\alpha$, the connection probability is flatter and alters are chosen uniformly at random.  The parameter $\alpha$ interpolates between heterogeneity and homogeneity in edge weights, even for ego networks with the same mean alter activity $t=\tau/k$ (\fref{fig:figure2}a; for a detailed model description see Materials and Methods [MM] and \srefsi{2}).

We solve the model analytically via a master equation for $p_a$ in the limit $\tau, k \to \infty$ (see MM and \srefsi{2} for derivation). By introducing the preferentiality parameter $\beta = t_r / \alpha_r$ with $t_r = t - a_0$ and $\alpha_r = \alpha + a_0$,  the activity distribution can be written as
\begin{equation}
\label{eq:actDist}
p_a = p_0 \frac{ a_r^{-1} }{ \mathrm{B}( a_r, \alpha_r ) } \left( 1 + \frac{1}{\beta} \right)^{-a_r},
\end{equation}
where $a_r = a - a_0$,  $p_0 = \left( 1 + \beta \right)^{-\alpha_r}$, and $\mathrm{B} (a_r, \alpha_r)$ is the Euler beta function. \eref{eq:actDist}  
fits to numerical simulations of the model very well, even for relatively low values of $\tau$ and $k$ (\fref{fig:figure2}b). The preferentiality parameter $\beta$,  the ratio between the average number of interactions in the ego network and the tendency of the ego and alters to interact preferentially, reveals a crossover in the behavior of the model, as signaled by the dispersion $d = \beta / (2 + \beta)$ (\fref{fig:figure2}c).  For large $\beta$,  dispersion increases (just like in the heterogeneous signatures of \fref{fig:figure1}) and $p_a$ takes the broad shape of a gamma distribution. When $\beta$ and $d$ are small, the activity distribution approaches a Poisson distribution and scales like a Gaussian in the limit of large $t_r$ (\fref{fig:figure2}d).

\begin{figure}[t]
\centering
\includegraphics[width=0.8\textwidth]{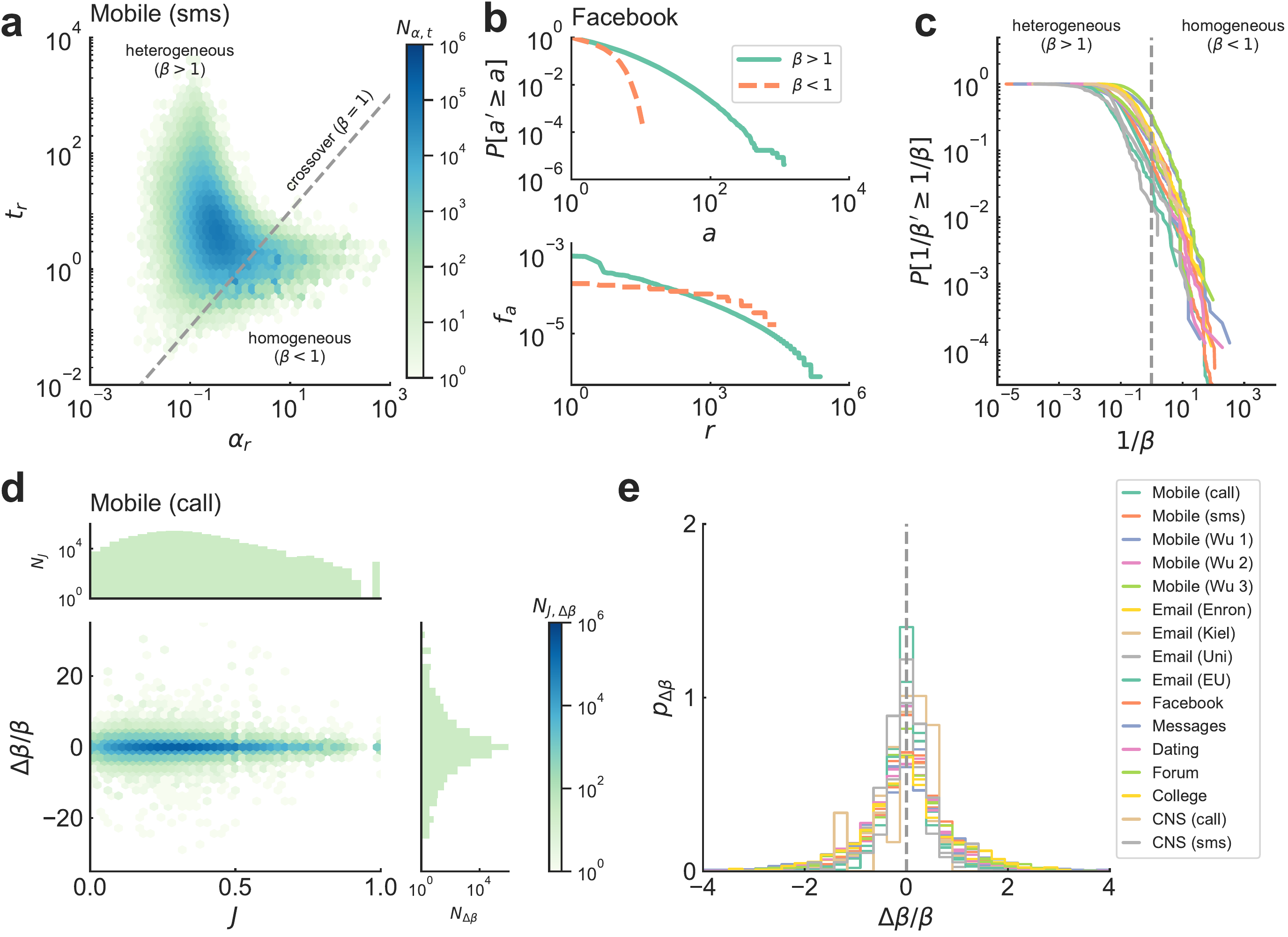}
\caption{\scriptsize {\bf Model reveals diversity and persistence of social signatures.}
{\bf (a)} Heat map of the number $N_{\alpha, t}$ of egos with given values of $\alpha_r = \alpha + a_0$ and $t_r = t - a_0$ in the Mobile (sms) dataset \cite{onnela2007analysis,onnela2007structure,karsai2011small,kivela2012multiscale,kovanen2013temporal,unicomb2018threshold,heydari2018multichannel} (data description in \srefsi{1}; all datasets in \frefsi{9}).  Most egos (93\%) have a heterogeneous social signature.  On the other side of the crossover $\beta=1$,  a few egos (7\%) have more homogeneous tie strengths (\trefsi{3}).  {\bf (b)} Complementary cumulative distribution $P[a' \geq a]$ of the number of alters having at least activity $a$ (top),  and fraction of events $f_a$ with alter in rank $r$ (bottom),  aggregated over all egos in the heterogeneous ($\beta > 1$) or homogeneous ($\beta < 1$) regime in the Facebook dataset~\cite{viswanath2009evolution,saramaki2015exploring}.  {\bf (c)} Complementary cumulative distribution $P[1/\beta' \geq 1/\beta]$ of rate $1 / \beta$,  estimated for 6.57M egos in 16 datasets of calls,  messaging,  and online interactions.  All systems show a diversity of social signatures,  with 66--99\% egos favouring a few of their alters,  and 1--34\% communicating homogeneously (\trefsi{3} and \frefsi{8}).  {\bf (d)} Number $N_{J,  \Delta \beta}$ of egos with given alter turnover $J$ and relative preferentiality change $\Delta \beta / \beta$ when estimating $\beta$ in two consecutive intervals of activity ($I_1$ and $I_2$, see \fref{fig:figure1} and \srefsi{3}), calculated for egos in the Mobile (call) dataset (all channels in \frefsi{10}).  We also show marginal number distributions of turnover ($N_J$) and relative preferentiality change ($N_{\Delta \beta}$).  Social signatures are persistent in time at the level of individuals,  regardless of alter turnover. {\bf (e)} Distribution $p_{\Delta \beta}$ of relative preferentiality change for all studied datasets.  Persistence of social signatures is systematic across communication channels.
}
\label{fig:figure3}
\end{figure}

Empirical ego networks have broadly distributed degree and minimum/mean alter activities for all communication channels studied (see \trefsi{1} and Fig. S1). With $k$, $a_0$, and $t$ fixed by the data,  \eref{eq:actDist} becomes a single-parameter model,  allowing us to derive maximum likelihood estimates for the preferentiality parameter $\beta$ in each ego network (\fref{fig:figure3}; see MM and \srefsi{3} for details on the fitting process). After performing a goodness-of-fit test~\cite{clauset2009power,morales2016generic,voitalov2019scale} with both Kolmogorov-Smirnov and Cramér-von Mises test statistics \cite{stephens1974edf},  we obtain $\beta$ estimates for $33-71\%$ of egos in each dataset, amounting to 6.57M individuals over 16 communication channels (\tsrefsi{2}{3}). Values of the preferentiality parameter, capturing the shape of the social signature of an ego, cover a wide region in the $(\alpha_r, t_r)$ space and accumulate around the crossover $\beta=1$ (\fref{fig:figure3}a; compare with \fref{fig:figure2}c; all datasets in \frefsi{9}).  By accumulating all alter activities over heterogeneous ($\beta > 1$) and homogeneous ($\beta < 1$) egos (\fref{fig:figure3}b),  both activity and activity-rank distributions have the same functional form as in \fref{fig:figure1}c, implying that the crossover value $d = 1/3$ predicted by the model is a more principled estimate of the boundary between regimes than the arbitrary threshold $d = \langle d \rangle$ (\fref{fig:figure1}d--e).

The heterogeneity of ego network tie strengths is well captured by the preferentiality parameter $\beta$, as it is a single number that encapsulates how each individual chooses which alters to interact with (cumulative advantage or effective random choice).  Our data and model show that this parameter is broadly distributed (66--99\% of ego networks in a dataset have heterogeneous and 1--34\% homogeneous signatures; see \trefsi{3}). Yet, the parameter has a similar functional shape in data representing different communication channels (\fref{fig:figure3}c).  To explore whether $\beta$ and the associated activity distribution $p_a$ are personal characteristics of each ego and not a product of random variation, we quantify its persistence by separating the communication activity of an ego into two consecutive intervals~\cite{saramaki2014persistence,aledavood2016channel,centellegher2017personality,heydari2018multichannel} (with the same number of events; see \fref{fig:figure1}a), fitting the model independently to each interval.  The difference $\Delta \beta$ in preferentiality, relative to $\beta$ for the whole observation period,  is very small for most egos (\fref{fig:figure3}d).  When separating individuals by alter turnover in their ego networks,  i.e. the Jaccard similarity coefficient $J$ between sets of alters in both intervals, the mean of $\Delta \beta$ remains close to zero even for egos with high network turnover ($J \sim 0$; for details see \srefsi{3} and \frefsi{10}).  The persistence of the preferentiality parameter,  found in all of our datasets regardless of communication channel (\fref{fig:figure3}e) and irrespectively of alter turnover,  shows that it indeed captures intrinsic individual differences in social behavior.

\section*{Discussion}

Our findings demonstrate that humans tend to build similar-looking personal networks on multiple online communication channels. The analysis of egocentric networks reveals a common heterogeneous pattern, in which a small group of alters receive a disproportionate amount of communication, yet substantial inter-individual variation is observed similarly across all datasets. To capture this pattern and its variation, we have developed a parsimonious and analytically tractable model of ego network evolution, which incorporates a preferentiality parameter specific to each ego. This parameter quantifies the degree of heterogeneity in an ego's personal network, reflecting the balance between two distinct mechanisms of tie reinforcement: cumulative advantage and random choice. Importantly, the distribution of fitted preferentiality parameter values characterizing individual social behavior is consistent across datasets from different channels, pointing to the presence of platform-independent universal patterns of communication.

This universality can be considered both expected and unexpected. In the case of people's ``real'' social networks, loosely defined as relationships that exist in the offline world, it is not surprising that their structure, characterized by a small number of close relationships, is reflected in online communication as well, such as through mobile phone calls. The cumulative advantage mechanism that drives the dispersion of tie strength can be simply thought to result from people putting more emphasis on their closest relationships. Generally, the heterogeneity of tie strengths in ego networks has been attributed to cognitive, temporal, and other constraints~\cite{miller1956magical,bernard1973social,dunbar1998social,gonccalves2011modeling,miritello2013time,lerman2016information}, and different personality traits~\cite{swickert2002extraversion, costa1992four} and their relative stability have been proposed as one possible reason for the persistent individual variation in this heterogeneity~\cite{centellegher2017personality}.

However, there is no a priori reason why the ego networks generated from work-related emails, dating website messages, or movie-related online forum discussions should exhibit similarities to those arising from mobile telephone communications. The nature of communication in these different contexts often pertains to a specific purpose and is limited to a subset of the ego's alters~\cite{PhysRevE.94.052319}, who may even only be represented by online aliases. Nevertheless, despite these differences, the overall pattern of heterogeneous tie strengths and the distribution of the preferentiality parameter, which captures inter-individual variability, are remarkably similar across all datasets. This raises questions as to the underlying mechanisms driving these similarities.

One possibility is that our brain is simply wired to consistently shape our social networks in similar ways, independent of the specific medium of communication~\cite{dunbar1998social,kock2005media}. Alternatively, the reason may lie in the mechanisms of tie strength reinforcement: cumulative advantage may arise, e.g., because we have already participated in an online conversation with someone and it is easier to continue interacting with the same alter. In other words, while the mechanism of cumulative advantage explains ego network tie strengths, it can arise because of different reasons: emotional closeness of real relationships  or the ease of repeated interactions in online communication with aliases. However, purely observational data such as those analyzed here cannot provide a clear answer, and thus further research is required.

An alternative perspective to consider is one in which all forms of social connections, whether they occur in-person or virtually, with actual people or pseudonymous entities, are integral components of an egocentric network that encompasses all relationships of an individual. Then, the various communication media can be viewed as distinct dimensions that reflect specific facets of this overarching network. Subnetworks associated with each communication channel are then shaped by the ego's channel preferences and may or may not contain the same alters (see, e.g., ~\cite{PhysRevE.94.052319}). It is conceivable that the cognitive and time constraints on personal networks act across the whole set of communication channels. Then, each individual has their own way of allocating their available communication activity on the different channels. The selection of a communication channel is known to affect the capacity to sustain emotionally  intense social relationships~\cite{vlahovic2012effects}, and it is plausible that channel-specific variations in an ego's preferentiality parameter may reflect their ability (or inability) to manage channel-specific constraints that impact effective social bonding. This offers additional insights into the debate surrounding competing theories such as media richness~\cite{daft1986organizational} and communication naturalness~\cite{kock2005media}. Given that the utilized datasets represent distinct populations, it is yet to be determined whether the preferentiality parameter of each individual displays similar or divergent values across different media. Recent research suggests that the values of the preferentiality parameter are similar at least for calls and text messages~\cite{heydari2018multichannel}, but it is not certain if this finding generalizes to other channels.

It is also notable that the value of the preferentiality parameter of each ego appears to be stable in time, even in the face of personal network turnover. This suggests that the parameter may reflect a persistent individual trait that influences the structure of egocentric networks on various channels. 
This interpretation raises important questions about the possible links between an ego's preferentiality parameter and their other personal characteristics, such as age, gender, and health. It is well established that the diversity of social relationships can serve as an indicator of increased longevity~\cite{holt2010social}, enhanced cognitive functioning during aging~\cite{fratiglioni2004active}, and greater resilience to disease~\cite{cohen2009can}.

Variation in the preferentiality parameter within a population may have also important consequences at the network level. Egocentric network tie strengths and their  variation are obviously related to the well-established heterogeneous distribution of tie strengths across the broader network (see, e.g.,~\cite{onnela2007analysis}). Moreover, if an ego's parameter value reflects a personal trait, it may also correlate with their network role.  For instance, in social media data, personality traits seem to correlate with the ability of an individual to increase their network size~\cite{quercia2012personality}, broker new relations between alters~\cite{staiano2012friends}, and participate in more communities~\cite{friggeri2012psychological}. 
Thus, a broad distribution of preferentiality parameter values among individuals may manifest as a macro-level network structure that reflects a broad array of roles and positions of individuals within the network. 
These observations highlight the potential for our findings to contribute to a broader understanding of the underlying mechanisms driving social network formation and individual behaviour.

\section*{Materials and Methods}

{\small \subsection*{Model of alter activity}

We consider a minimal ego network dynamics where individuals allocate interactions via cumulative advantage and a tunable amount of random choice (for details see \srefsi{2}). At initial event time $\tau_0 = k a_0$ with $k$ the degree of the ego network, all alters have minimal activity $a_0$. At any time $\tau \geq \tau_0$, the probability that an alter with activity $a$ becomes active at time $\tau+1$ is
\begin{equation}
\label{eq:connProbBeta}
\pi_a = \frac{a_r / t_r + \beta^{-1}}{ k ( 1 + \beta^{-1} ) },
\end{equation}
with $a_r = a - a_0$, $t_r = t - a_0$, and $t = \tau / k$ the mean alter activity. The preferentiality parameter $\beta = t_r / \alpha_r$ (with $\alpha_r = \alpha + a_0$ and $\alpha$ a tunable parameter) interpolates between two regimes: random alter choice ($\beta \to 0$ and $\pi_a \sim 1/k$), and preferential alter selection ($\beta \to \infty$ and $\pi_a \sim a_r / \tau_r$ with $\tau_r = \tau - \tau_0$).

The model can be treated analytically in the limit $\tau, k \to \infty$ with constant $t$  (\srefsi{2}). The probability $p_a$ that a randomly chosen alter has activity $a$ follows the master equation
\begin{equation}
\label{eq:masterEq}
d_t p_a = \frac{1}{t + \alpha} \big[ (a - 1 + \alpha) p_{a - 1} - (a + \alpha) p_a \big],
\end{equation}
with initial condition $p_a(a_0) = \delta_{a,a_0}$ and $d_t$ the derivative with respect to $t$. By introducing the probability generating function $g(z,t) = \sum_a p_a z^a$, \eref{eq:masterEq} reduces to
\begin{equation}
\label{eq:PFG_PDE}
\partial_t g = \frac{z - 1}{t + \alpha} \big( z \partial_z g + \alpha g \big),
\end{equation}
a partial differential equation with initial condition $g(z, a_0) = z^{a_0}$.  Via the method of characteristics, $g$ takes the explicit form
\begin{equation}
\label{eq:PGF}
g(z, t) = z^{a_0} \left[ z + (1 - z) \left( 1 + \beta \right) \right]^{-\alpha_r},
\end{equation}
from which we obtain the activity distribution $p_a$ in \eref{eq:actDist} iteratively by taking partial derivatives of $g$ with respect to $z$. The distribution $p_a$ has mean $t$ and variance $\sigma^2 = t_r (1 + \beta)$, leading to the dispersion index $d = \beta / (2 + \beta)$.

}

{\small \subsection*{Fitting data and model}

We derive maximum likelihood estimates of the model parameter for empirical ego networks with degree $k$, minimum/maximum alter activity $a_0$ and $a_m$, and total/mean alter activity $\tau = \sum_i a_i$ and $t = \tau/k$ (for details see \srefsi{3}). Assuming that the $k$ alter activities $\{ a_i \}$ are independent and identically distributed random variables following $p_a$ in the model, the likelihood $L_{\alpha}$ that the sample $\{ a_i \}$ is generated by \eref{eq:actDist} for given $\alpha$ follows
\begin{equation}
\label{eq:logDeriv}
d_{\alpha} \ln L_{\alpha} = k \left[ F_{\alpha} - \ln ( 1 + \beta ) \right],
\end{equation}
where $F_{\alpha} = \frac{1}{k} \sum_i [ \psi ( a_r + \alpha_r ) - \psi ( \alpha_r ) ]$ is an average over all observed relative activities $a_r = a_i - a_0$ of the digamma function $\psi(\alpha) = d_{\alpha} \Gamma(\alpha) / \Gamma(\alpha)$, i.e. the logarithmic derivative of the gamma function $\Gamma(\alpha)$.  The $\alpha$ value that maximizes $L_{\alpha}$ is given implicitly by
\begin{equation}
\label{eq:alphaTrascEq}
\alpha_r = \frac{ t_r }{ e^{ F_{\alpha} } - 1 },
\end{equation}
or, equivalently,  by $\beta = e^{ F_{\alpha} } - 1$.

A goodness-of-fit test allows us to quantify how plausible is the hypothesis that the empirical data is drawn from the model activity distribution in \eref{eq:actDist} (\srefsi{3}).  We measure goodness of fit via the standard Kolmogorov-Smirnov statistic
\begin{equation}
\label{eq:KSstat}
D = \max_{a_0 \leq a \leq a_m} | \Delta P_a |,
\end{equation}
that is, the largest magnitude of the difference  $\Delta P_a (t) = P_{\mathrm{data}}[a' \leq a] - P_a(t)$ between the cumulative distribution of alter activity in data, $P_{\mathrm{data}}[a' \leq a]$, and that of the fitted model,  $P_a(t) = \sum_{a'=a_0}^a p_{a'}(t)$, across all activities $a \in [a_0, a_m]$. We check the robustness of our results with three other measures from the Cramér-von Mises family of test statistics (for details see \srefsi{3}). 

Given the sample $\{ a_i \}$, we compute the estimate $\alpha$ numerically from \eref{eq:alphaTrascEq} and the statistic $D$ from \eref{eq:KSstat}, where the model activity distribution follows \eref{eq:actDist}. From the model we generate $n_{\mathrm{sim}}=2500$ simulated activity samples $\{ a_i \}_{\mathrm{sim}}$. For each simulated sample, we find its own estimate $\alpha_{\mathrm{sim}}$ and the corresponding statistic $D_{\mathrm{sim}}$. Then, the fraction of simulated statistics $D_{\mathrm{sim}}$ larger than the data statistic $D$ is the $p$-value associated with the goodness-of-fit test, according to $D$. If the $p$-value is large enough ($p > 0.1$ with 0.1 an arbitrary significance threshold), we do not rule out the hypothesis that our activity model emulates the empirical data, and we consider that the ego network has a measurable preferentiality parameter  $\beta$. We aim at obtaining large $p$-values (rather than small),  since we want to keep the assumption that the model is a good description of the observed data (rather than reject it). Our goodness-of-fit test shows that $33-71\%$ of all considered ego networks are well described by the model (or up to $42-88\%$ for other test statistics; see \trefsi{2}).
}

{\small \subsubsection*{Data and code availability}
Code to reproduce the results of the paper is publicly available at \url{https://github.com/iniguezg/Farsignatures}. For data availability see \srefsi{1}. Non-public data is available from the authors upon reasonable request.

\subsubsection*{Acknowledgments}
G.I.  thanks Tiina Näsi for valuable suggestions.  G.I. and J.K. acknowledge support from AFOSR (Grant No. FA8655-20-1-7020), project EU H2020 Humane AI-net (Grant No. 952026), and CHIST-ERA project SAI (Grant No. FWF I 5205-N). J.K. acknowledges support from European Union’s Horizon 2020 research and innovation programme under grant agreement ERC No 810115 - DYNASNET. We acknowledge the computational resources provided by the Aalto Science--IT project. The study was part of the NetResilience consortium funded by the Strategic Research Council at the Academy of Finland (grant numbers 345188 and 345183).}

\subsubsection*{Author contributions}
G.I., S.H., J.K., and J.S. conceived, designed, and developed the study.  G.I. and S.H. analyzed empirical data.  G.I. derived analytical results and performed numerical simulations and model fitting.  G.I., S.H., J.K., and J.S. wrote the paper.

\subsubsection*{Competing interest statement}
All authors declare no competing interest.

{\footnotesize \bibliographystyle{ieeetr}
\bibliography{ms}
}

\end{document}

\begin{center}
{\LARGE Supplementary Information for}\\[0.7cm]
{\Large \textbf{Universal patterns in egocentric communication networks}}\\[0.5cm]

{\large G. Iñiguez$^*$, S. Heydari, J. Kertész, J. Saramäki$^*$}\\[0.7cm]
{\small $^*$Corresponding author email: iniguezg@ceu.edu, saramaki@aalto.fi}\\[1cm]
\end{center}

\addtocontents{toc}{\protect\setstretch{0.1}}
\tableofcontents

\section{Communication data}
\label{sec:data}

\begin{figure}[t]
\centering
\includegraphics[width=0.8\textwidth]{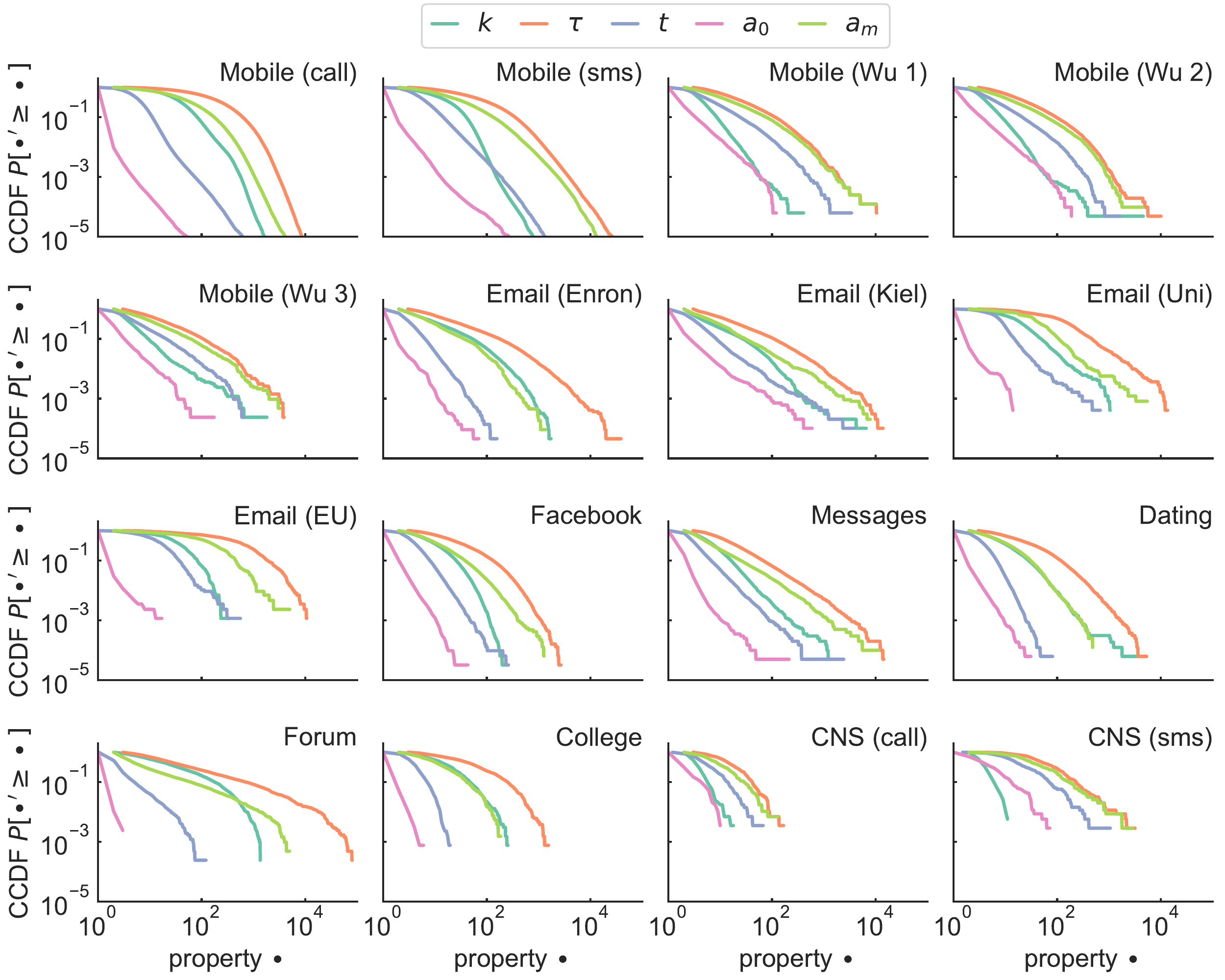}
\caption{\small {\bf Basic properties of communication datasets.} Complementary cumulative distribution functions (CCDFs) $P[\bullet' \geq \bullet]$ for several properties $\bullet$ of ego networks in each dataset: degree $k$ (number of alters of an ego), strength $\tau$ (number of events involving an ego), mean alter activity $t$ (average number of events per alter of an ego), minimum activity $a_0$ (minimum number of events with the same alter), and maximum activity $a_m$ (maximum number of events with the same alter). All properties are heterogeneously distributed across egos and alters, with some differences between datasets.}
\label{fig:dataCCDF}
\end{figure}

We analyze several datasets of social interactions between individuals from a wide range of studies in the temporal networks literature (\tref{tab:datasets} and \fref{fig:dataCCDF}). Each dataset includes a time-ordered set of communication events between anonymized individuals $i$ and $j$ (according to hashed timestamps).  For each dataset, we construct temporal ego networks for each individual so that the network for ego $i$ contains all events where $i$ participates. Therefore, each event connecting nodes $i$ and $j$ appears both in the ego network where $j$ is an alter of ego $i$, and in the ego network where $i$ is an alter of ego $j$ (except otherwise explicitly stated in \sref{sec:dataList}). \tref{tab:datasets} lists basic properties of all datasets considered, starting with the system size $N_u$ (unfiltered number of egos) and number of events $V$ (all distinct contact events between egos and alters). We only consider egos with any level of heterogeneous alter activity, i.e.  with mean alter activity $t$ larger than the minimum across its alters ($t > a_0$), leading to a reduced system size $N$ (filtered number of egos). \tref{tab:datasets} includes several properties of the filtered datasets: average degree $\langle k \rangle$ (mean number of alters per ego), average strength $\langle \tau \rangle$ (mean number of events per ego), average mean alter activity $\langle t \rangle$ (mean number of events per alter per ego), and average minimum/maximum alter activity $\langle a_0 \rangle$ and $\langle a_m \rangle$ (mean of lowest/highest alter activity per ego). We briefly describe below each dataset considered, including references to detailed studies and locations of publicly available data.

\subsection{List of datasets}
\label{sec:dataList}

\begin{table}[!ht]
\small
\noindent\makebox[\textwidth]{ \begin{tabular}{l l r r | r r r r r r}
\toprule
Dataset & Event & $N_u$ & $V$ & $N$ & $\langle k \rangle$ & $\langle \tau \rangle$ & $\langle t \rangle$ & $\langle a_0 \rangle$ & $\langle a_m \rangle$ \\
\midrule
Mobile (call)~\cite{onnela2007analysis,onnela2007structure,karsai2011small,kivela2012multiscale,kovanen2013temporal,unicomb2018threshold,heydari2018multichannel} & Phone calls & 5994967 & 1342862618 & 5431921 & 38.84 & 246.29 & 5.91 & 1.02 & 69.20 \\
Mobile (sms)~\cite{onnela2007analysis,onnela2007structure,karsai2011small,kivela2012multiscale,kovanen2013temporal,unicomb2018threshold,heydari2018multichannel} & Short messages & 5387745 & 613751054 & 4233187 & 16.95 & 143.30 & 7.49 & 1.15 & 60.68 \\
Mobile (Wu 1)~\cite{wu2010evidence} & Short messages & 44090 & 544817 & 16050 & 4.55 & 52.93 & 12.74 & 1.84 & 38.10 \\
Mobile (Wu 2)~\cite{wu2010evidence} & Short messages & 71042 & 636629 & 20534 & 4.71 & 43.86 & 10.66 & 1.91 & 29.86 \\
Mobile (Wu 3)~\cite{wu2010evidence} & Short messages & 14273 & 140611 & 4215 & 6.27 & 52.72 & 10.66 & 1.79 & 33.29 \\
Email (Enron)~\cite{klimt2004enron,kunegis2013konect} & Emails & 86978 & 1134990 & 21984 & 22.52 & 96.43 & 3.26 & 1.15 & 16.27 \\
Email (Kiel)~\cite{ebel2002scale,saramaki2015exploring} & Emails & 57189 & 431864 & 9842 & 13.05 & 65.68 & 5.99 & 1.79 & 25.86 \\
Email (Uni)~\cite{eckmann2004entropy,saramaki2015exploring} & Emails & 3188 & 308730 & 2456 & 25.49 & 250.10 & 9.14 & 1.12 & 61.18 \\
Email (EU)~\cite{leskovec2007graph,paranjape2017motifs} & Emails & 986 & 332334 & 866 & 36.92 & 766.98 & 18.22 & 1.08 & 194.64 \\
Facebook~\cite{viswanath2009evolution,saramaki2015exploring} & Online messages & 45813 & 854612 & 31429 & 11.04 & 53.22 & 4.08 & 1.17 & 17.07 \\
Messages~\cite{said2010social,karimi2014structural,saramaki2015exploring} & Online messages & 35623 & 478015 & 20252 & 8.37 & 45.14 & 3.84 & 1.23 & 17.85 \\
Dating~\cite{holme2004structure,saramaki2015exploring} & Online messages & 28972 & 430826 & 16239 & 13.05 & 51.44 & 3.44 & 1.11 & 13.29 \\
Forum~\cite{said2010social,karimi2014structural,saramaki2015exploring} & Online messages & 7084 & 1428493 & 4122 & 65.22 & 691.15 & 2.83 & 1.01 & 57.41 \\
College~\cite{opsahl2009clustering,panzarasa2009patterns} & Online messages & 1899 & 59835 & 1303 & 20.48 & 90.90 & 3.62 & 1.07 & 17.22 \\
CNS (call)~\cite{stopczynski2014measuring,sapiezynski2019interaction} & Phone calls & 525 & 3234 & 285 & 3.25 & 19.00 & 6.25 & 1.76 & 12.47 \\
CNS (sms)~\cite{stopczynski2014measuring,sapiezynski2019interaction} & Short messages & 568 & 24333 & 347 & 3.36 & 114.73 & 33.89 & 5.66 & 86.50 \\
\bottomrule
\end{tabular}}
\caption{
\small {\bf Datasets used in this study}.
Characteristics of the available datasets, starting with system size $N_u$ (unfiltered number of egos) and number of events $V$ (all communication events between egos and alters). We only consider egos with mean alter activity larger than its minimum ($t > a_0$), leading to a system of size $N$ (filtered number of egos) with the following properties: average degree $\langle k \rangle$ (mean number of alters per ego), average strength $\langle \tau \rangle$ (mean number of events per ego), average mean alter activity $\langle t \rangle$ (mean number of events per alter per ego), and average minimum/maximum alter activity $\langle a_0 \rangle$ and $\langle a_m \rangle$ (mean of lowest/highest alter activity per ego). We include references to detailed studies of each dataset and locations of publicly available data.
}
\label{tab:datasets}
\end{table}

\paragraph{Phone calls (Mobile call \& sms).} Mobile phone records dataset consisting of time-stamped communication logs between anonymized users, originally introduced in \cite{onnela2007analysis}.  Data covers logs on outgoing communication of approximately 20\% of the population of an undisclosed European country and spans a 6-months period in 2007. We have filtered out self-communication events (e.g., users messaging themselves) and records made by people under so-called `family contracts' with the operator company (indicating that several individuals might have used the same phone line).  After filtering, the dataset includes more than 5 million users,  1.3 billion calls, and 613 million short messages. Unlike the rest of the datasets below, this data is directional, meaning ego networks consist only of outgoing communication events. Data is not publicly available, but has been extensively studied in the literature (see, for example, \cite{onnela2007analysis,onnela2007structure,karsai2011small,kivela2012multiscale,kovanen2013temporal,unicomb2018threshold,heydari2018multichannel}).

\paragraph{Short messages (Wu 1, 2 \& 3).} Dataset from a mobile phone operator including three charging accountant bills from three companies (denoted 1, 2, and 3) over a 1-month period. Each event comprises a sender mobile phone number, a recipient mobile phone number (both anonymized), and a hashed timestamp with a precision of 1 second~\cite{wu2010evidence}. Data is publicly available in the Supplementary Information of~\cite{wu2010evidence}.

\paragraph{Emails (Enron).} Dataset of email communication from the Enron corporation during 1999--2003, which was made public as a result of legal action by the Federal Energy Regulatory Commission in the US.  A subset of the corpus including $200, 399$ messages sent between 158 users was originally studied in 2004~\cite{klimt2004enron}.  In 2015 this corpus was corrected and published in raw form~\cite{enron2015}. Data we use comes from the Koblenz network collection~\cite{kunegis2013konect} and corresponds to $1,148,072$ emails between $87,273$ addresses, both inside and outside Enron. After filtering out events with equal sender and recipient, we obtain the slightly lower values of $N_u$ and $V$ in \tref{tab:datasets}. Data is publicly available at \url{http://konect.cc/networks/enron/}.

\paragraph{Emails (Kiel).} Dataset of log files of email server at Kiel University, recording source and destination of every email from or to a student account over a period of 112 days~\cite{ebel2002scale}. Data has also been analyzed in terms of temporal greedy walks in~\cite{saramaki2015exploring} (see \sref{sec:dataAcks} for data acknowledgments).

\paragraph{Emails (Uni).} Dataset of log files of one of the main mail servers at an unnamed university, comprising email messages sent during a period of 83 days and connecting $\sim$ 10,000 users~\cite{eckmann2004entropy}. Data was reduced to the internal mail within the institution, leaving a set of 3,188 users interchanging 309,125 messages. The dataset has also been analyzed in terms of temporal greedy walks in~\cite{saramaki2015exploring}. The value of $V$ in~\cite{eckmann2004entropy} slightly differs when calculated directly from available data (see \tref{tab:datasets} and \sref{sec:dataAcks} for data acknowledgments).

\paragraph{Emails (EU).}  Dataset of email communication in a large European research institution from October 2003 to May 2005, comprising 3,038,531 messages between 287,755  addreses~\cite{leskovec2007graph,paranjape2017motifs}. After focusing only on institution members and the emails sent between them, values of $N_u$ and $V$ decrease to those in \tref{tab:datasets}. Data is publicly available at the SNAP repository in \url{https://snap.stanford.edu/data/email-Eu-core-temporal.html}.

\paragraph{Online messages (Facebook).} Dataset on both friendship relationships and interactions for a large subset of the Facebook New Orleans social network, comprising over 60,000 anonymized users and over 800,000 logged interactions (wall posts) between users in a period of two years~\cite{viswanath2009evolution}. Facebook links were crawled during December 29th, 2008 and January 3rd, 2009, starting from a single user and visiting friends with a breadth-first-search algorithm. Wall posts were then crawled between January 20th, 2009 and January 22nd, 2009 for all previously detected users. Wall post data spans from September 26th, 2006 to January 22nd, 2009. The dataset has also been analyzed in terms of temporal greedy walks in~\cite{saramaki2015exploring}. Data is publicly available at: \url{http://socialnetworks.mpi-sws.org/data-wosn2009.html}. Values of $N_u$ and $V$ in~\cite{viswanath2009evolution} differ when calculated directly from available data (see \tref{tab:datasets}).

\paragraph{Online messages (Messages \& Forum).} Dataset from the social movie recommendation community Filmtipset (Sweden's largest and available since 2000), consisting of time-stamped communications (contact events) between 36,492 users during 7 years~\cite{karimi2014structural}. Available data corresponds to a user-to-user messaging channel where each user can send text messages to another user privately and only one user at a time (Messages), and an open forum where users comment on posts of other users, as many as are willing to participate (Forum). The dataset was originally studied in~\cite{said2010social}, and has also been analyzed in terms of temporal greedy walks in~\cite{saramaki2015exploring} (see \sref{sec:dataAcks} for data acknowledgments).

\paragraph{Online messages (Dating).} Dataset from pussokram.com, a Swedish online community primarily intended for romantic communication and targeted at adolescents and young adults, consisting of all activity during 512 days from 13 February 2001 to 10 July 2002 among roughly 30,000 users~\cite{holme2004structure}. Time-stamped contact events between users follow 4 modes of communication: private intra-community emails, guest book signing, friendship requests (`flirts'), and friendships. Data has also been analyzed in terms of temporal greedy walks in~\cite{saramaki2015exploring} (see \sref{sec:dataAcks} for data acknowledgments).

\paragraph{Online messages (College).} Dataset of private messages sent on a Facebook-like online social network for students at the University of California, Irvine, from April to October 2004,  where users could search the network for others and then initiate conversations based on their profile information~\cite{opsahl2009clustering,panzarasa2009patterns}. Data includes the 1,899 students that sent or received at least one message on the site, comprising 59,835 online messages over 20,296 directed ties between these users.  The dataset is hosted by Tore Opsahl at \url{https://toreopsahl.com/datasets/#online_social_network} and is also publicly available from the SNAP repository at \url{https://snap.stanford.edu/data/CollegeMsg.html}.

\paragraph{Copenhagen Networks Study (CNS call \& sms)} Dataset of multi-channel, phone-enabled social interactions from the Copenhagen Networks Study (CNS)~\cite{stopczynski2014measuring,sapiezynski2019interaction}. The original study includes activity of roughly 1,000 individuals during 2012-2013 via Bluetooth interactions, calls, and messages~\cite{stopczynski2014measuring}. Data used here is a selected portion of the full dataset as described in~\cite{sapiezynski2019interaction}. The selected dataset includes call and short message logs between individuals, with data on timestamps of the call/message, anonymized user IDs, and call duration. We disregard missed calls, making the dataset smaller from the one in~\cite{sapiezynski2019interaction}. Data is publicly available via {\it figshare} in~\cite{sapiezynski2019copenhagen}.

\subsubsection{Data acknowledgments}
\label{sec:dataAcks}

We are grateful for data provision to Albert-László Barabási (Mobile call \& sms) and Petter Holme (Email Kiel \& Uni, Forum, Messages, Dating). We acknowledge the Department of Computer Science of Aalto University for access to processed versions of the non-public datasets used here (Mobile call \& sms, Email Kiel \& Uni, Forum, Messages, Dating).

\subsection{Ego network properties, activity dispersion and connection kernel}
\label{sec:dataProps}

\begin{figure}[t]
\centering
\includegraphics[width=0.8\textwidth]{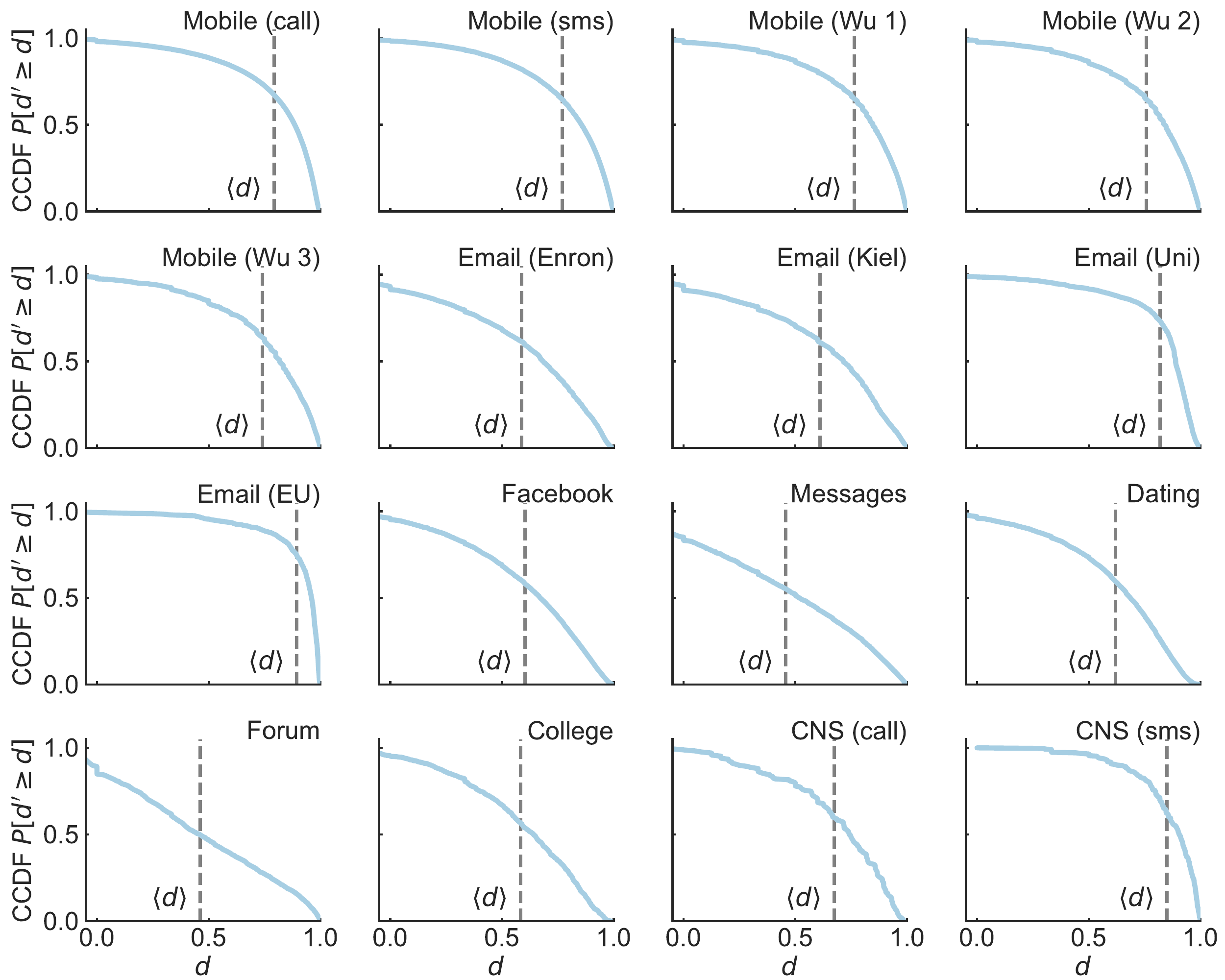}
\caption{\small {\bf Dispersion in communication activity.} Complementary cumulative distribution function (CCDF) $P[d' \geq d]$ of the number of egos having at least dispersion index $d$ in all considered datasets,  calculated only for ego networks with more than 10 events, i.e. $\tau>10$.  The average dispersion $\langle d \rangle$ is displayed as a dashed line. Communication channels show broad variation in how egos allocate activity among alters.}
\label{fig:dispersion}
\end{figure}

In all considered datasets of communication, ego networks have heterogeneous structures and patterns of activity. For each ego network we measure the degree $k$, strength $\tau$, mean alter activity $t$, minimum alter activity $a_0$, and maximum alter activity $a_m$, and then see how these measure vary across egos. All properties show broad tails in their corresponding complementary cumulative distribution functions (CCDFs), the probability $P[\bullet' \geq \bullet]$ that an ego has property $\bullet'$ larger than a given value $\bullet$ (see \fref{fig:dataCCDF}). 

In order to measure the variability in communication patterns between egos and alters (the heterogeneity of tie strengths in an ego network), we focus on the alter activity distribution $p_a$,  the probability that a randomly chosen alter has activity $a$.  Following~\cite{goh2008burstiness}, we quantify the spread of $p_a$ via the variance-to-mean ratio $\sigma_r^2 / \mu_r$ by defining the dispersion index
\begin{equation}
\label{eq:dispIndex}
d = \frac{\sigma_r^2 / \mu_r - 1}{\sigma_r^2 / \mu_r + 1} = \frac{\sigma_r^2 - \mu_r}{\sigma_r^2 + \mu_r}
\end{equation}
for each ego network in a dataset, where $\mu_r = t_r = t - a_0$ is the mean alter activity relative to the minimum $a_0$,  and $\sigma^2_r = \sigma^2$ is the (location-invariant) variance of alter activity. The CCDF $P[d' \geq d]$ (fraction of egos having at least dispersion index $d$) varies smoothly with $d$ in all systems, meaning there are egos with both narrow ($d \sim 0$) and broad ($d \sim 1$) alter activity distributions (\fref{fig:dispersion}).

\begin{figure}[t]
\centering
\includegraphics[width=0.8\textwidth]{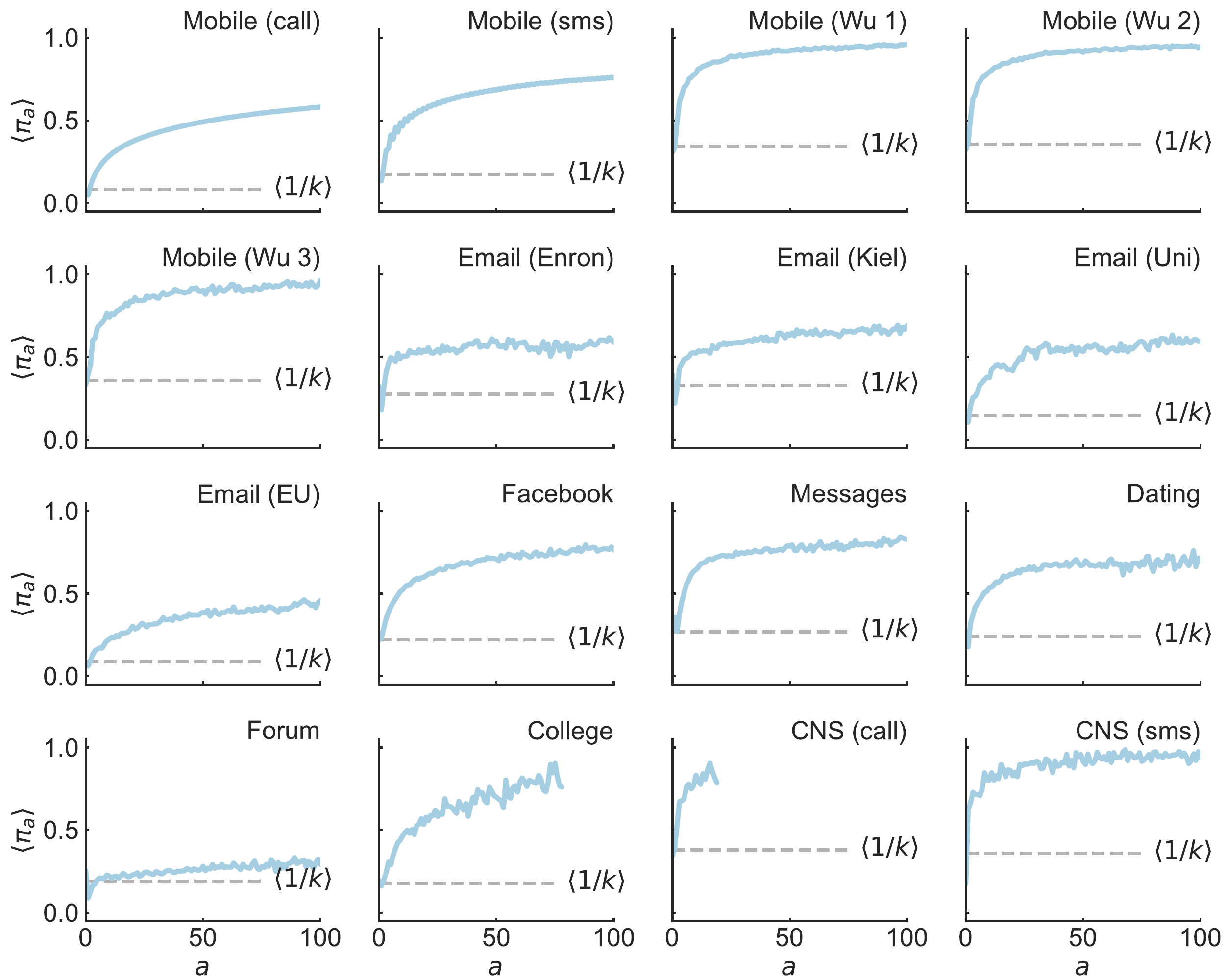}
\caption{\small {\bf Connection kernel in communication.} Probability $\langle \pi_a \rangle$ that an alter with current activity $a$ communicates once more with the ego, averaged over time and subsets of at least 50 egos with degree $k \geq 2$ for each $a$ value, shown here for all considered datasets. The dashed line corresponds to the average baseline $\langle \pi_a \rangle = \langle 1/k \rangle$ when communication events are distributed randomly. The growth of the connection kernel $\langle \pi_a \rangle$ with activity indicates cumulative advantage, where alters with high prior activity receive more communication. }
\label{fig:kernel}
\end{figure}

We also calculate the connection kernel $\pi_a$, the probability that an alter with current activity $a$ communicates once more with the ego.  When averaged over time and large enough subsets of ego-alter pairs with activity $a$ at some point in time, the average connection kernel $\langle \pi_a \rangle$ increases with alter activity (\fref{fig:kernel}). Apart from low values of $a$,  $\langle \pi_a \rangle$ is larger than the average baseline $\langle 1/k \rangle$ (communication events are distributed uniformly at random among alters) and increases faster than linear or roughly linearly for all datasets, depending on the activity range. This behavior indicates cumulative advantage: alters with high prior activity are more likely to communicate with the ego later on in time.

\clearpage

\section{Model of alter activity}
\label{sec:model}

Consider a social ego network made up of one central individual (the ego) and its $k$ acquaintances (the alters), where a tie between ego and alter represents communication activity between individuals (i.e. calls/messages or online interactions). At a discrete event time $\tau$ (starting from $\tau_0$ up to the length of the observation window), each alter $i = 1, \ldots, k$ has an activity score $a_i(\tau)$ counting the number of times the ego and alter $i$ have communicated until time $\tau$. We take as initial condition $a_i( \tau_0 ) = a_0$ for all $i$, meaning that all alters have the same initial activity $a_0 \geq 0$, i.e. the minimum activity observed across alters. At each time $\tau$ of the model dynamics, a single alter $i$ communicates with the ego, such that $a_i( \tau + 1 ) = a_i(\tau) + 1$. Taking $\tau_0 = k a_0$, we ensure that event time is equal to the sum of all communication events in the ego network, i.e. $\tau = \sum_i a_i$ is the total communication activity. Scores are thus bounded by the growing interval $a_i(\tau) \in [\tau_0, \tau]$.

We consider a cumulative-advantage dynamics (similar to Price's model~\cite{price1965networks,price1976general,newman2018networks}) tuned by a parameter $\alpha$: the probability $\pi_a(\tau)$ that an alter with previous activity $a_i(\tau) = a$ is active at time $\tau + 1$ is proportional to its past number of communications,
\begin{equation}
\label{eq:CumAdvDef}
\pi_a(\tau) = \frac{a + \alpha}{\sum_j [ a_j(\tau) + \alpha ]} = \frac{a + \alpha}{ \tau + k \alpha }.
\end{equation}
The connection kernel in \eref{eq:CumAdvDef} is well defined (at any time $\tau \geq \tau_0$) for any $\alpha$ larger than its minimum value $\alpha_0 = -a_0$, so we can also tune the model by the relative parameter $\alpha_r = \alpha  - \alpha_0 = \alpha + a_0 > 0$. Similarly, we define the relative alter activity $a_r = a - a_0$, relative event time $\tau_r = \tau - \tau_0 = \tau - k a_0$, and the relative mean alter activity $t_r = t - t_0 = t - a_0$ with $t = \tau / k$. Introducing the preferentiality parameter
\begin{equation}
\label{eq:betaScale}
\beta = \frac{t_r}{\alpha_r} = \frac{t - a_0}{\alpha + a_0}
\end{equation}
allow us to rewrite \eref{eq:CumAdvDef} as
\begin{equation}
\label{eq:CumAdvDefGamma}
\pi_a = \frac{a_r / t_r + \beta^{-1}}{ k ( 1 + \beta^{-1} ) }.
\end{equation}
As we will see in \sref{sec:MasterEqDeriv}, the scale $\beta$ (or, alternatively, the rate $\beta^{-1}$) quantifies a crossover between regimes of behavior in alter activity. For $\beta \ll 1$ (i.e. $\alpha \to \infty$ for fixed $t$ and $a_0$), we have $\pi_a = 1/k$ for any $a_r$ and communication events are spread uniformly at random among alters. For $\beta \gg 1$ (i.e. $\alpha \to \alpha_0$ for fixed $t$ and $a_0$), the probability of communication is roughly proportional to activity, $\pi_a \to a_r / \tau_r$. In this way, the preferentiality parameter $\beta$ interpolates between a {\it homogeneous regime} where communication in the ego network is uniformly random ($\beta < 1$), and a {\it heterogeneous regime} where activity is driven by cumulative advantage ($\beta > 1$), with a crossover at $\beta = 1$ ($\alpha_r = t_r$).

\subsection{Master equation for activity dynamics}
\label{sec:MasterEqDeriv}

We treat our model analytically by solving a master equation for the activity dynamics in the limit of large total alter activity $\tau \to \infty$ and large number of alters $k \to \infty$, such that the mean alter activity $t = \tau / k$ is kept constant. We denote by $p_a(\tau)$ the time-dependent probability that an alter chosen uniformly at random has activity $a$ at time $\tau$., i.e. the alter activity distribution When a new communication event happens at time $\tau + 1$, with probability $\pi_a$ the group of $k p_a$ alters with activity $a$ loses one alter (since the alter's activity increases to $a+1$). With probability $\pi_{a - 1}$ the group also wins one alter from the group of $k p_{a - 1}$ alters with activity $a - 1$ (since the alter's activity increases to $a$). The master equation for $p_a$ is
\begin{equation}
\label{eq:MasterEqDisc}
p_a( \tau + 1 ) - p_a(\tau) = \pi_{a - 1}(\tau) p_{a - 1}(\tau) - \pi_a(\tau) p_a(\tau),
\end{equation}
with initial condition $p_a (\tau_0) = \delta_{a, a_0}$, and $p_{a} \equiv 0$ for $a < a_0$.

Taking the limit $\tau, k \to \infty$ (with $dt = 1/k \to 0$) and rescaling time to the fixed mean alter activity $t = \tau dt$, we can rewrite \eref{eq:MasterEqDisc}  as a continuous master equation for the alter activity distribution $p_a(t)$,
\begin{equation}
\label{eq:MasterEqCont}
d_t p_a = \frac{1}{t + \alpha} \big[ (a - 1 + \alpha) p_{a - 1} - (a + \alpha) p_a \big],
\end{equation}
with $d_t$ the derivative with respect to $t$. The initial time is $ t_0 = \tau_0 / k = a_0$, so the initial condition of \eref{eq:MasterEqCont} is $p_a(t_0) = \delta_{a, a_0}$, with $p_a \equiv 0$ for $a < a_0$. We solve \eref{eq:MasterEqCont} within a generating function formalism. We introduce the probability generating function (PGF) $g(z, t)$ associated to $p_a$,
\begin{equation}
\label{eq:ProbGenFunc}
g(z, t) = \sum_{a = 0}^{\infty} p_a(t) z^a.
\end{equation}
which returns the probability $p_a$ by computing the $a$-th partial derivative with respect to $z$, $p_a(t) = \partial^a_z g (0, t) / a!$.
Summing up over $a$ in \eref{eq:MasterEqCont} and manipulating dummy indices, we obtain a partial differential equation (PDE) for $g$,
\begin{equation}
\label{eq:PFG_PDE}
\partial_t g = \frac{z - 1}{t + \alpha} \big( z \partial_z g + \alpha g \big),
\end{equation}
with initial condition $g(z, t_0) = z^{a_0}$.

The linear PDE in \eref{eq:PFG_PDE} can be solved with the method of characteristics. By introducing an auxiliary variable $s$, solving \eref{eq:PFG_PDE} is equivalent to solving the system of ordinary differential (Lagrange-Charpit) equations for $t \equiv t(s)$, $z \equiv z(s)$ and $g \equiv g(s)$,
\begin{equation}
\label{eq:MethCharODEs}
\begin{cases}
d_s t = t + \alpha, & t(0) = a_0,\\
d_s z = z(1 - z), & z(0) = z_0, \\
d_s g = \alpha (z - 1) g, & g(0) = z_0^{a_0}.
\end{cases}
\end{equation}
Using the solutions of \eref{eq:MethCharODEs} to substitute $s$ and $z_0$, we obtain an explicit expression for the PGF,
\begin{equation}
\label{eq:PGF}
g(z, t) = z^{a_0} \left[ z + (1 - z) \frac{t + \alpha}{a_0 + \alpha} \right]^{- (a_0 + \alpha)} = z^{a_0} \left[ z + (1 - z) \left( 1 + \beta \right) \right]^{-\alpha_r},
\end{equation}
where we use the preferentiality parameter $\beta = t_r / \alpha_r$ (with $\alpha_r = \alpha + a_0$ and $t_r = t - a_0$).

\begin{figure}[t]
\centering
\includegraphics[width=0.8\textwidth]{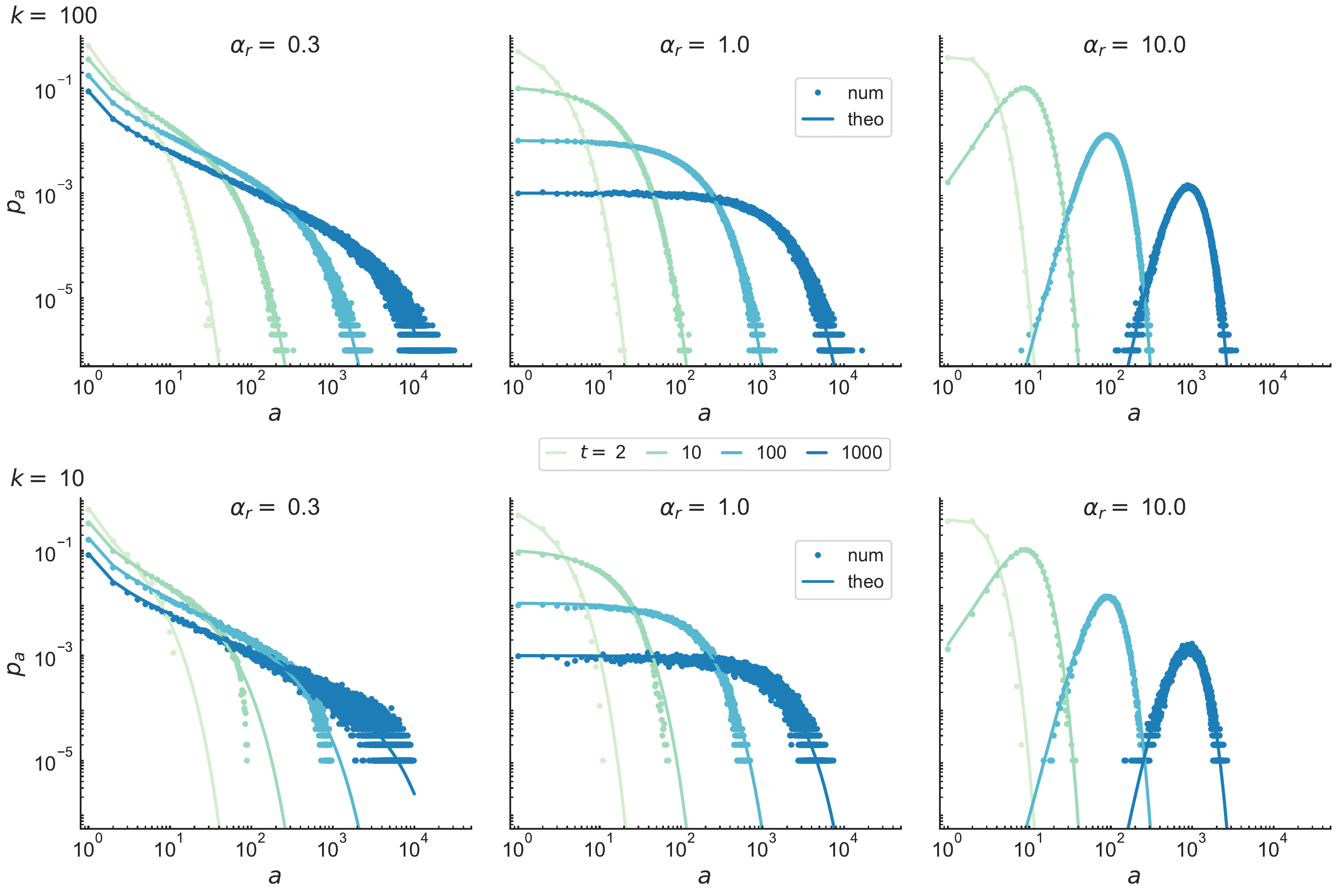}
\caption{\small {\bf Simple model of alter activity.} Probability $p_a(t)$ of a randomly selected alter having activity $a$ at time $t$, as a function of $a = a_r + a_0$ for varying $t = t_r + a_0$ and varying $\alpha_r = \alpha + a_0$, for fixed $a_0 = 1$ and $k = 10, 100$ (bottom/top rows), in both numerical simulations of the model [\eref{eq:CumAdvDef}; dots] and its analytical solution [\eref{eq:ActDistExp}; lines]. For given $t$ and $a_0$, the time evolution of $p_a$ [as defined by \eref{eq:MasterEqCont}] reaches an $\alpha_r$-dependent asymptotic shape. {\bf (right)} When $\alpha_r \to \infty$ ($\alpha \to \infty$), $p_a$ converges to a Poisson distribution with mean and variance $t_r$ [\eref{eq:ActDistLargeAlpha}]. {\bf (left)} When $\alpha_r \to 0$ ($\alpha \to -a_0$), $p_a$ approaches a gamma distribution with shape $\alpha_r$ and scale $\beta = t_r / \alpha_r$ [\eref{eq:ActDistSmallAlpha}]. {\bf (middle)} When $\alpha_r = 1$, the exponent of the power-law decay in the gamma distribution is $\alpha_r - 1 = 0$, so the activity distribution has a plateau of $a$ values with relatively constant $p_a$ that grows with $t$. \eref{eq:ActDistExp} approximates numerical simulations very well, but fails at the tail for sufficiently low $k$. Simulations are averaged over $10^4$ realizations.}
\label{fig:activity_num_approx}
\end{figure}

Equating terms between \eref{eq:ProbGenFunc} and the Mclaurin series of \eref{eq:PGF} lets us calculate the alter activity distribution $p_a$ explicitly by calculating partial derivatives of the PGF $g$ with respect to $z$. After some algebra and by using the preferentiality parameter $\beta$ we obtain
\begin{equation}
\label{eq:ActDistExp}
p_a(t) = p_0 \frac{ a_r^{-1} }{ \mathrm{B}( a_r, \alpha_r ) } \left( 1 + \frac{1}{\beta} \right)^{-a_r}
\end{equation}
for $a_r > 0$ ($a > a_0$), with
\begin{equation}
\label{eq:normConstant}
p_0 = p_{a_0}(t) = \left( 1 + \beta \right)^{-\alpha_r}
\end{equation}
for $a_r = 0$ ($a = a_0$) and $p_a \equiv 0$ for $a_r < 0$ ($a < a_0$). For consistency, $p_a(t) = \delta_{a, a_0}$ for $t_r = 0$ ($t = a_0$). In \eref{eq:ActDistExp}, $\mathrm{B} (a_r, \alpha_r) = \Gamma (a_r) \Gamma (\alpha_r) / \Gamma ( a_r + \alpha_r )$ is the Euler beta function, with $\Gamma (a_r) = (a_r - 1)!$  the gamma function.

The $n$-th raw moment of $p_a$ can also be computed from \eref{eq:PGF} as $m^{(n)} = (z \partial_z)^n g |_{z=1}$, leading to the mean $\mu = t$ (in consistence with the definition of the model) and variance $\sigma^2 = t_r (1 + \beta)$. Changing variables from $a$ to the relative alter activity $a_r = a - a_0$, we obtain the relative mean $\mu_r = t_r$ and variance $\sigma^2_r = \sigma^2$ (since variance is location-invariant). This allows us to write the dispersion index $d$ of \eref{eq:dispIndex} in terms of $\beta$ as
\begin{equation}
\label{eq:burstParam}
d = \frac{\sigma_r^2 - \mu_r}{\sigma_r^2 + \mu_r} = \frac{\beta}{2 + \beta}.
\end{equation}

\eref{eq:ActDistExp} has an intuitive behaviour as a function of the relative alter activity $a_r$, mean alter activity $t_r$, cumulative-advantage parameter $\alpha_r$, and the minimum alter activity $a_0$ (\fref{fig:activity_num_approx}). Even if the derivation of \eref{eq:ActDistExp} assumes $\tau, k \to \infty$ for fixed $t$, its functional form agrees very well with numerical simulations of the dynamical rule in \eref{eq:CumAdvDef} for degree as low as $k = 100$ (\fref{fig:activity_num_approx} upper row), with some disagreement in the tail of the activity distribution for even lower $k = 10$ due to finite-size effects (\fref{fig:activity_num_approx} lower row). The first factor in \eref{eq:ActDistExp}, $p_0$, shows that the fraction of alters with minimum activity decreases as time goes by with a decay regulated by $\alpha_r$. The second factor, $a_r^{-1} / \mathrm{B}( a_r, \alpha_r )$, is roughly a power law for intermediate values of activity $a_r$ with exponent regulated by $\alpha_r$. The third factor is an exponential cutoff for large activity $a_r$ at the scale $\beta$ that moves to the right as time $t_r$ increases. As we will see below, the behaviour of the activity distribution $p_a(t)$ is even more apparent by approximating \eref{eq:ActDistExp} in the heterogeneous ($\beta > 1$) and homogeneous ($\beta < 1$) regimes by either a gamma or Poisson distribution. Since these distributions have their own scaling form, $\beta$ parametrises a crossover between regimes in terms of the scaling of the activity distribution.

\subsubsection{Heterogeneous regime \texorpdfstring{($\beta > 1$)}{}: Alter activity is gamma-distributed}
\label{sec:gammaRegime}

We explore the limit $\alpha_r \to 0$ ($\alpha \to -a_0$ for fixed $a_0$) by considering a large activity $a_r \gg 0$ ($a \gg a_0$, i.e. the tail of the activity distribution) for small but fixed $\alpha_r$. Then, the beta function behaves as $\mathrm{B} (a_r, \alpha_r) \simeq \Gamma (\alpha_r) a_r^{-\alpha_r}$ for given $\alpha_r$. The condition $\beta > 1$ leads to the approximations $(1 + \beta)^{-\alpha_r} \simeq \beta^{-\alpha_r}$ and $(1 + 1 / \beta)^{-a_r} \simeq e^{-a_r / \beta}$ (from a 1st-order Taylor expansion of the exponential). Inserting into \eref{eq:ActDistExp} we obtain
\begin{equation}
\label{eq:ActDistSmallAlpha}
p_a(t) = \frac{ 1 }{ \beta^{\alpha_r} \Gamma(\alpha_r) } a_r^{\alpha_r - 1} e^{ -a_r / \beta }, \quad \alpha_r \to 0,
\end{equation}
a gamma distribution with shape $\alpha_r$ and scale $\beta$. Then, the relative alter activity $a_r$ has mean $t_r$ and variance $t_r \beta$. Consistently, $\sigma_r^2 \to \infty$ as $\beta \to \infty$, implying a dispersion $d \to 1$ [see \eref{eq:burstParam}]. In the heterogeneous regime where alters communicate with the ego with probability proportional to their previous activity, the activity distribution $p_a(t)$ has power-law behaviour with exponent $\alpha_r - 1$ and an exponential cutoff regulated by the scale $\beta$ (see, e.g., \fref{fig:activity_num_approx} left).

\begin{figure}[t]
\centering
\includegraphics[width=0.8\textwidth]{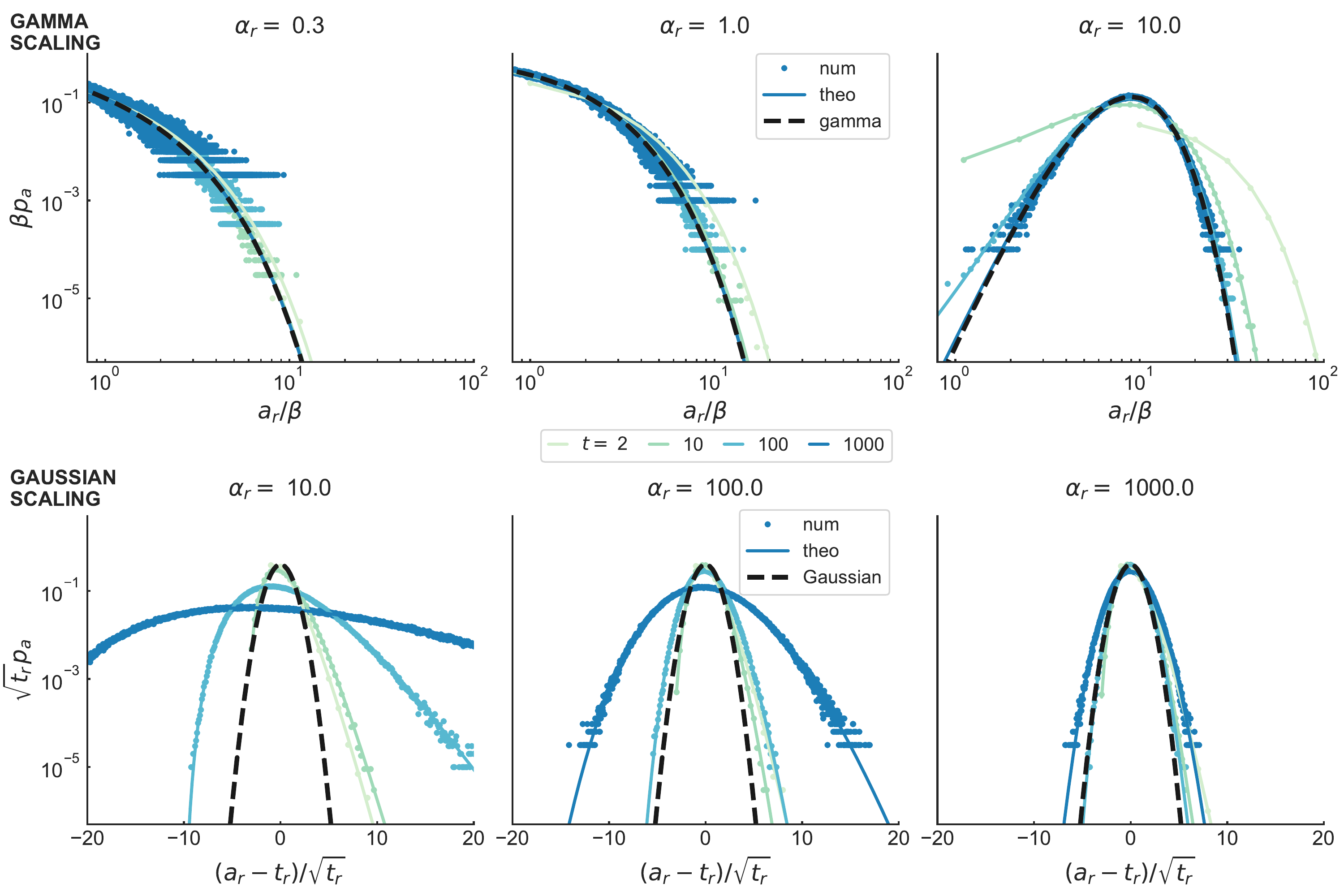}
\caption{\small {\bf Crossover in scaling of alter activity.} Probability $p_a(t)$ of a randomly selected alter having activity $a$ at time $t$, as a function of $a_r = a - a_0$ for varying $t = t_r + a_0$ and varying $\alpha_r = \alpha + a_0$, for fixed $a_0 = 1$ and $k = 100$, in both numerical simulations of the model [\eref{eq:CumAdvDef}; dots] and its analytical solution [\eref{eq:ActDistExp}; lines]. {\bf (top)} By plotting $\beta p_a$ vs. $a_r / \beta$ for varying $t_r$ and fixed $\alpha_r$, curves collapse to the standard gamma distribution [\eref{eq:gammaScaling}; dashed lines]. This $\alpha_r$-dependent, gamma scaling is valid in the heterogeneous regime $\beta > 1$, with $\beta = t_r / \alpha_r$ the scale parameter of the gamma distribution in \eref{eq:ActDistSmallAlpha}. Though only asymptotically correct, gamma scaling is a good approximation even at the crossover $\beta = 1$. {\bf (bottom)} By plotting $\sqrt{ t_r } p_a$ vs. $( a_r - t_r ) / \sqrt{ t_r }$ for varying $t_r$, curves collapse to the standard Gaussian distribution [\eref{eq:GaussianScaling}; dashed lines]. This Gaussian scaling is valid in the homogeneous regime $\beta < 1$ and becomes asymptotically more accurate with $t_r$. Simulations are averaged over $10^4$ realizations.}
\label{fig:activity_scaling}
\end{figure}

The moment-generating function of the gamma distribution shows that \eref{eq:ActDistSmallAlpha} has exponential scaling. Plugging the rescaled activity $a_r' = a_r / \beta$ into \eref{eq:ActDistSmallAlpha} leads to
\begin{equation}
\label{eq:gammaScaling}
\beta p_a(t) = \frac{ 1 }{ \Gamma(\alpha_r) } \left( \frac{ a_r }{ \beta } \right)^{\alpha_r - 1} e^{ -a_r / \beta },
\end{equation}
the standard gamma distribution (with shape $\alpha_r$ and scale 1). In a plot of $\beta p_a$ vs. $a_r / \beta$ for varying $t_r$ and fixed $\alpha_r$, all curves collapse to the standard form of \eref{eq:gammaScaling} (\fref{fig:activity_scaling} top row). The gamma distribution (and its scaling property) is a very good approximation of the activity distribution even for relatively low activity $a_r$, as long as we are in the heterogeneous regime of $\beta > 1$ (for example, gamma scaling fails in the top right plot of \fref{fig:activity_scaling} for $t = 2$ and $\alpha_r = 10$ since $\beta = 0.1$). The gamma scaling shape in the heterogeneous regime depends on $\alpha_r$, but remains a good approximation of the activity distribution even at the crossover $\beta = 1$.

\begin{figure}[t]
\centering
\begin{subfigure}[b]{0.4\textwidth}
\includegraphics[width=\textwidth]{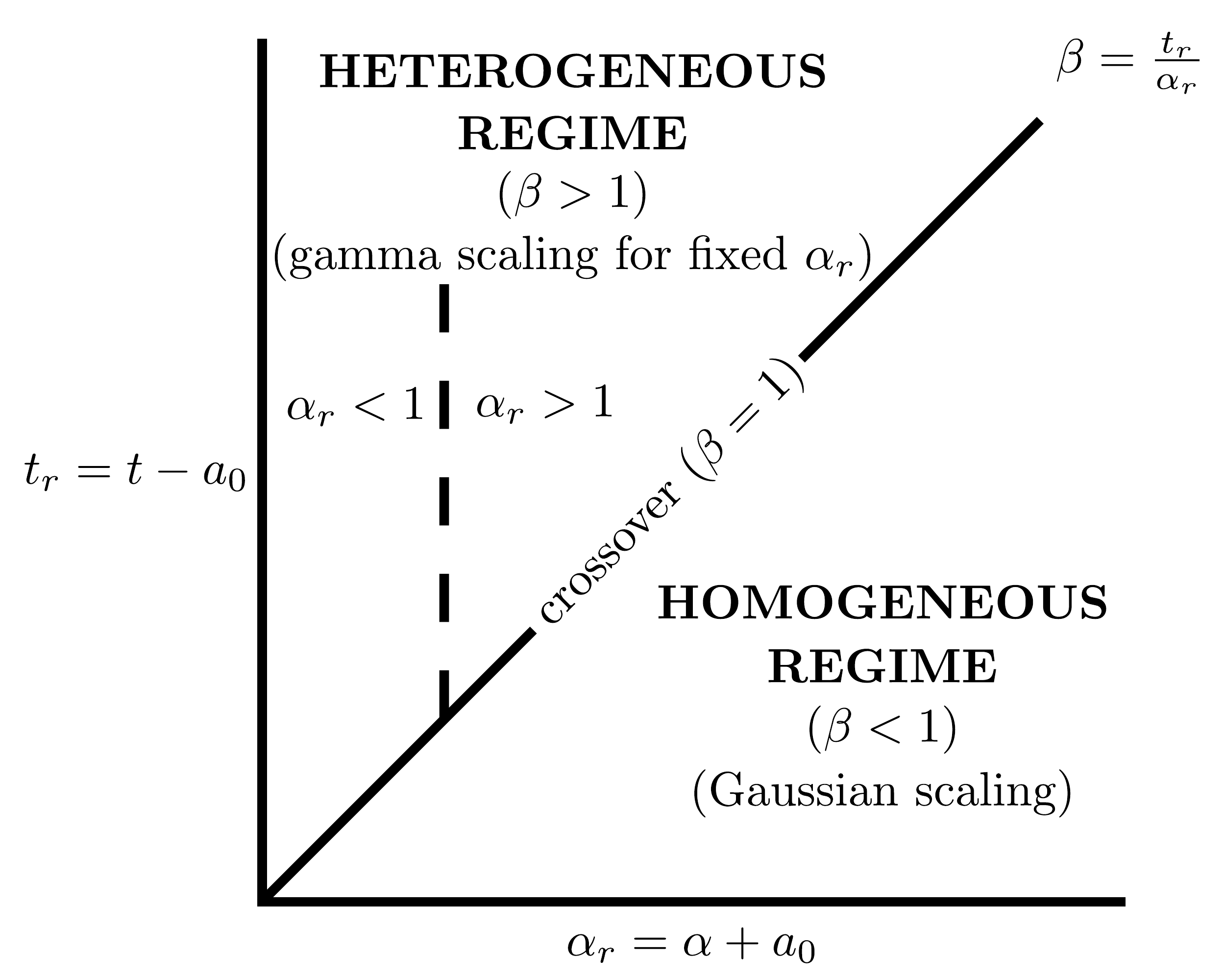}
\end{subfigure}
\\[0.1cm]
\begin{subfigure}[b]{0.8\textwidth}
\includegraphics[width=\textwidth]{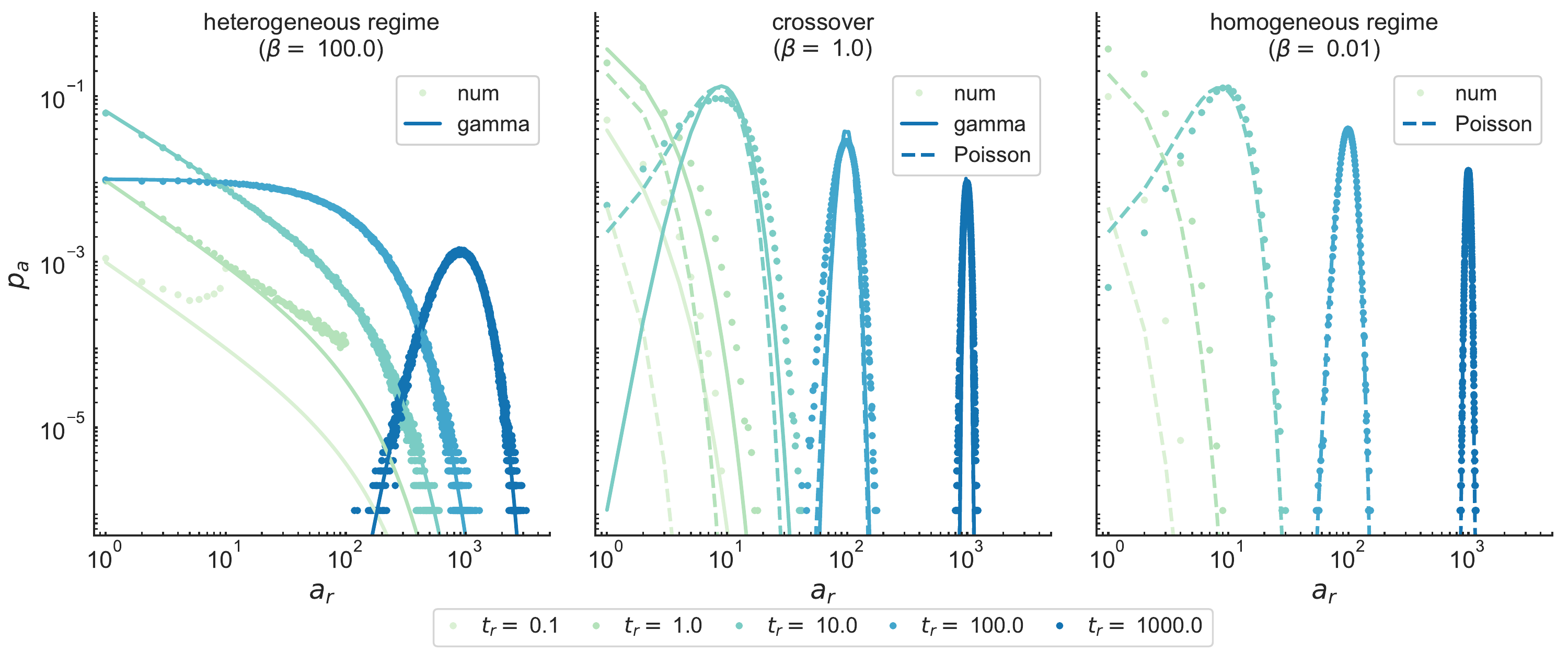}
\end{subfigure}
\caption{\small {\bf Scaling regimes in alter activity.} {\bf (top)} Phase diagram in ($\alpha_r, t_r$)-space, showcasing the scaling regimes in alter activity of the model defined by \eref{eq:CumAdvDef}. {\bf (bottom)} Probability $p_a(t)$ of a randomly selected alter having activity $a$ at time $t$, as a function of $a_r = a - a_0$ for varying $t_r = t - a_0$ and varying $\alpha_r = \alpha + a_0$ such that $\beta = t_r / \alpha_r$ is constant, for fixed $a_0 = 1$ and $k = 100$, in both numerical simulations of the model [\eref{eq:CumAdvDef}; dots] and the gamma [\eref{eq:ActDistSmallAlpha}] and Poisson [\eref{eq:ActDistLargeAlpha}] approximations. When $\beta < 1$ (right), activity is homogeneously distributed across alters and $p_a(t)$ is Poissonian with asymptotic Gaussian scaling. When $\beta >  1$ (left), $p_a(t)$ has gamma scaling for varying $t_r$ and fixed $\alpha_r$. For $\alpha_r < 1$ a few alters accumulate most activity, and for $\alpha_r > 1$ alter activity is more homogeneously distributed. Regimes are separated by a crossover at $\beta = 1$ where both gamma and Gaussian scaling forms fail slightly.}
\label{fig:activity_regimes}
\end{figure}

\subsubsection{Homogeneous regime \texorpdfstring{($\beta < 1$)}{}: Alter activity is Poisson-distributed}
\label{sec:randomRegime}

In the limit $\alpha_r \to \infty$ ($\alpha \to \infty$ for fixed $a_0$) where alters communicate with the ego uniformly at random, the beta function behaves as $\mathrm{B} (a_r, \alpha_r) \simeq \Gamma (a_r) \alpha_r^{-a_r}$ for given $a_r$. For $\beta < 1$, we can approximate $( 1 + \beta )^{-\alpha_r} \simeq e^{-t_r}$ (from a 1st-order Taylor expansion of the exponential) and $( 1 + 1 / \beta )^{-a_r} \simeq \beta^{a_r}$. Then, the activity distribution converges to a Poisson distribution with mean and variance $t_r$,
\begin{equation}
\label{eq:ActDistLargeAlpha}
p_a(t) = \frac{ t_r^{a_r} e^{-t_r} }{ a_r! }, \quad \alpha_r \to \infty.
\end{equation}
Thus, for large $\alpha_r$, the decay of the activity distribution $p_a$ is exponential and independent of $\alpha_r$ (\fref{fig:activity_num_approx} right panel). Since $\sigma_r^2 \to t_r$ as $\beta \to 0$, this limit consistently recovers a dispersion $d \to 0$  [\eref{eq:burstParam}].

The scaling behaviour of the Poissonian activity distribution in \eref{eq:ActDistLargeAlpha} is apparent from exploring the limit $t_r \to \infty$ (i.e. large mean alter activity $t$ for fixed $a_0$). Using Stirling's approximation of the gamma function, $a_r! = \Gamma( a_r + 1 ) \simeq \sqrt{ 2 \pi a_r } e^{ -a_r } a_r^{a_r}$, and assuming that $p_a$ only takes significant values close to $a_r = t_r$, the activity distribution approaches a Gaussian distribution with mean $t_r$ and standard deviation $\sqrt{t_r}$,
\begin{equation}
\label{eq:GaussianScaling}
p_a(t) = \frac{ 1 }{ \sqrt{ 2 \pi t_r} } e^{ -\frac{ ( a_r - t_r )^2 }{ 2t_r } }, \quad t_r \to \infty. 
\end{equation}
When plotting $\sqrt{ t_r } p_a$ vs. $( a_r - t_r ) / \sqrt{ t_r }$ for varying $t_r$, all curves collapse to the standard Gaussian distribution with mean 0 and standard deviation 1 (\fref{fig:activity_scaling} bottom row). The Poisson distribution (and its Gaussian scaling property) is a very good approximation of the activity distribution even for relatively low $t_r$, as long as we are in the homogeneous regime of $\beta < 1$ (for example, Gaussian scaling fails in the bottom center plot of \fref{fig:activity_scaling} for $t = 1000$ and $\alpha_r = 100$ since $\beta = 10$). Note that the asymptotic Gaussian scaling shape in the homogeneous regime is independent of $\alpha_r$, but it converges slowly as we increase $t_r$.

Overall, the model of alter activity in social ego networks defined by the dynamical rule in \eref{eq:CumAdvDef} has two regimes of behavior in $(\alpha_r, t_r)$-space regulated by the preferentiality parameter $\beta = t_r / \alpha_r$ (\fref{fig:activity_regimes} top). In the homogeneous regime of $\beta < 1$, the activity distribution $p_a(t)$ is asymptotically Poissonian with Gaussian scaling for increasing $\alpha_r$ and $t_r$, meaning that the ego spreads events homogeneously across its alters, with no strong dependence on the particular value of $\alpha_r$. In the heterogeneous regime of $\beta > 1$, $p_a(t)$ is well approximated by a gamma distribution, meaning that the activity distributions for varying $t_r$ and fixed $\alpha_r$ scale together (vertical lines in the upper triangle of \fref{fig:activity_regimes} top). The activity distribution is either monotonically decreasing for $\alpha_r < 1$ (a few alters accumulate most events) or has a broad peak around $t_r$ for $\alpha_r > 1$ (alter activity is a bit more homogeneous). When $\beta = 1$, both gamma and Gaussian scaling forms fail to reproduce numerical simulations slightly (\fref{fig:activity_regimes} bottom), implying a crossover between the heterogeneous and homogeneous regimes. The largest disagreement with simulations occurs for really low $t_r$ and $\alpha_r$, due to the assumption $\tau, k \to \infty$ in the derivation of \sref{sec:MasterEqDeriv}.

\clearpage

\section{Fitting data and model}
\label{sec:fitting}

\subsection{Derivation of maximum likelihood estimates}
\label{sec:MLEderiv}

The analytical derivation of the activity distribution $p_a(t)$ in the limit $\tau, k \to \infty$ for fixed mean alter activity $t$ and given minimum alter activity $a_0$ (see \sref{sec:MasterEqDeriv}) allows us to write explicitly a maximum likelihood estimate (MLE) for the model parameter $\alpha$ associated with an empirical observation of alter activities. Take a finite ego network of $k$ alters observed during a window of length $t_r = t - a_0$, where each alter $i$ has $a_i$ communication events with the ego ($i = 1, \ldots, k$), and the least active alter has $a_0$ events~\footnote{Note that the results of \sref{sec:MasterEqDeriv} are strictly valid only in the limit $\tau, k \to \infty$, but we use them to obtain MLEs in empirical ego networks with finite (and heterogeneously distributed) degree $k$.}. Since in empirical data we recognize alters when there is at least a single communication event with the ego, we always measure $a_0 > 0$. The total alter activity is $\tau = \sum_i a_i$ and the empirical mean alter activity is $t = \tau / k$, so the only free parameter in the model is $\alpha$. Assuming that alter activities are independent and identically distributed random variables following the activity distribution $p_a$ in \esref{eq:ActDistExp}{eq:normConstant}, the likelihood $L_{\alpha}$ that the sample $\{ a_i \}$ is generated by $p_a$ for a certain $\alpha$ value is given by the product
\begin{equation}
\label{eq:likelihoodFunc}
L_{\alpha} = \prod _{i = 1}^k p_{a_i} (t) = p_0^k \prod_{a_r \neq 0} \frac{ a_r^{-1} }{ \mathrm{B}( a_r, \alpha_r ) } \left( 1 + \frac{1}{\beta} \right)^{-a_r},
\end{equation}
with $p_0 = (1 + \beta)^{-\alpha_r}$ from \eref{eq:normConstant}, where we use the relative quantities $a_r = a_i - a_0$, $\alpha_r = \alpha + a_0$, and the preferentiality parameter $\beta = t_r / \alpha_r$. Like in \eref{eq:CumAdvDef}, we have the constraints $\alpha_r > 0$ and $a_r \geq 0$ for all $i$.

Taking the natural logarithm of \eref{eq:likelihoodFunc} and its derivative with respect to $\alpha$ leads, after some algebra, to
\begin{equation}
\label{eq:logDeriv}
d_{\alpha} \ln L_{\alpha} = k \left[ F_{\alpha} - \ln ( 1 + \beta ) \right]
\end{equation}
where $F_{\alpha} = \frac{1}{k} \sum_i [ \psi ( a_r + \alpha_r ) - \psi ( \alpha_r ) ]$ is an average over all observed relative activities $a_r = a_i - a_0$ of the digamma function $\psi(\alpha) = d_{\alpha} \Gamma(\alpha) / \Gamma(\alpha)$, i.e. the logarithmic derivative of the gamma function $\Gamma(\alpha)$. The MLE $\hat{\alpha}$ that maximizes $L_{\alpha}$ with respect to $\alpha$ (or, alternatively, the MLE $\hat{\beta} = t_r / \hat{\alpha}_r$ with $\hat{\alpha}_r = \hat{\alpha} + a_0$), is given implicitly by
\begin{equation}
\label{eq:MLEmax}
d_{\alpha} \ln L_{\alpha} |_{\alpha = \hat{\alpha}} = 0.
\end{equation}
\esref{eq:logDeriv}{eq:MLEmax} lead to a trascendental equation for the estimated cumulative-advantage parameter $\hat{\alpha}$,
\begin{equation}
\label{eq:alphaTrascEq}
\hat{\alpha}_r = \frac{ t_r }{ e^{ F_{\hat{\alpha}} } - 1 },
\end{equation}
or, equivalently, $\hat{\beta} = e^{ F_{\hat{\alpha}} } - 1$ for the estimated preferentiality parameter $\hat{\beta}$. Observe that \eref{eq:alphaTrascEq} is valid only if $a_i > a_0$ for at least some alter $i$, so that the empirical activity distribution is different from the initial condition in the model [$p_a(t_0) = \delta_{a, a_0}$]. When all alters have the same activity [i.e. $a_i = a_0$ for all $i$, so $a_r = t_r = \beta = F_{\alpha} = 0$ for nonzero $\alpha_r$], \eref{eq:MLEmax} is trivially valid for any $\alpha$ and we cannot use the MLE procedure to find an optimal value.  This is consistent with the filtering condition $t > a_0$ for empirical communication data introduced in \sref{sec:data}, implying that we can potentially find MLEs for all considered ego networks.

\begin{figure}[t]
\centering
\includegraphics[width=\textwidth]{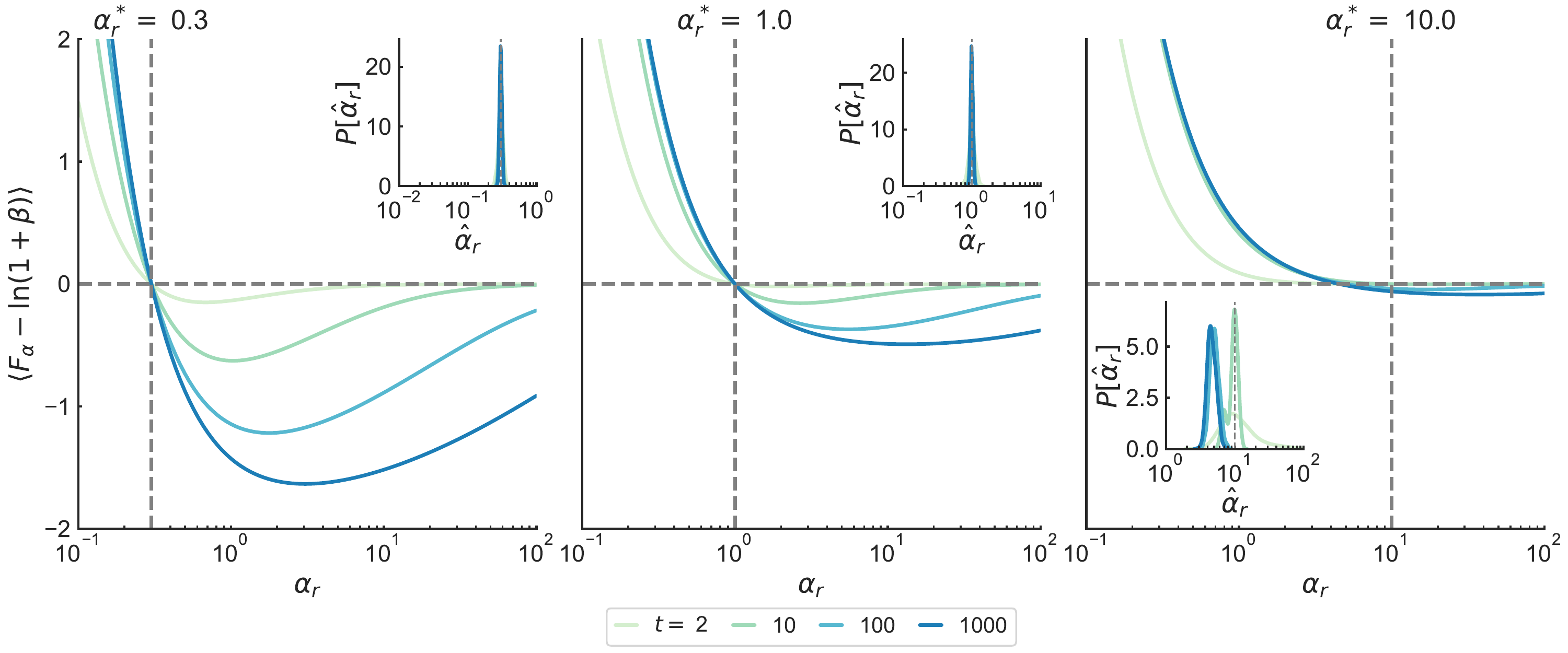}
\caption{\small {\bf Consistency of maximum likelihood estimation.} Numerical simulations of alter activity [according to \eref{eq:CumAdvDef}] for several target values $\alpha_r^*$ of the model parameter and varying $t = t_r + a_0$, with $a_0 = 1$ and $k = 10^3$. 
The quantity $\langle F_{\alpha} - \ln ( 1 + \beta ) \rangle$ as a function of $\alpha_r$ (an average over $10^3$ simulations),  has a root $\hat{\alpha}_r$ somewhere close to $\alpha_r^*$, in accordance with \eref{eq:alphaTrascEq}. Insets show the kernel density estimate $P[\hat{\alpha}_r]$ of the MLE $\hat{\alpha}_r$ [as computed numerically from \eref{eq:alphaTrascEq}] over all simulations,  which is centered around $\alpha_r^*$ for most parameter values.  The MLE procedure recovers the target value $\alpha_r^*$ and is thus consistent,  apart from a systematic underestimation for large $t$ in the homogeneous regime.}
\label{fig:figure_MLEfit}
\end{figure}

We can verify the accuracy of the MLE method by solving \eref{eq:alphaTrascEq} graphically and numerically for synthetic activity distributions $p_a(t)$ coming from many stochastic realizations of the model rule in \eref{eq:CumAdvDef} (\fref{fig:figure_MLEfit}). We obtain samples $\{ a_i \}$ from numerical simulations of the model for given $k$, $t$, and $a_0$, for several target values $\alpha_r^*$ of the model parameter. Then we plot the average $\langle F_{\alpha} - \ln ( 1 + \beta ) \rangle$ over all realizations as a function of $\alpha_r$ to graphically locate a root, and we also compute $\hat{\alpha}_r$ numerically from \eref{eq:alphaTrascEq}, which follows the distribution $P[\hat{\alpha}_r]$ over all realizations (inset in \fref{fig:figure_MLEfit}). The MLE procedure is quite accurate ($\hat{\alpha}_r \sim \alpha_r^*$) and thus consistent in the heterogeneous and crossover regimes. In the homogeneous regime, however, $\hat{\alpha}_r$ systematically underestimates the target value $\alpha_r^*$ for large $t$, where the functional form of $\langle F_{\alpha} - \ln ( 1 + \beta ) \rangle$ does not depend much on $t$ anymore. It is instructive to see this MLE bias in light of the scaling property of the activity distribution. In the homogeneous regime, alter activities asymptotically scale like a Gaussian regardless of the value of $\alpha$, so even relatively large errors in estimating $\alpha$ lead to the same scaling form. In the heterogeneous regime, where activities have an $\alpha$-dependent gamma scaling, less bias means we can estimate the scaling form of empirical data more accurately.

\subsection{Goodness-of-fit test}
\label{sec:GOFtest}

The MLE $\hat{\alpha}$ given implicitly by \eref{eq:alphaTrascEq} is the value of $\alpha$ maximizing the likelihood that the activity model of \sref{sec:model} produces a given empirical activity distribution. In addition, we need a goodness-of-fit (GOF) test quantifying how plausible is the hypothesis that the empirical data is drawn from the theoretical activity distribution $p_a(t)$ in \eref{eq:ActDistExp}. Following \cite{clauset2009power}, we measure goodness of fit by means of the distance between the activity distributions in model and data. (We have previously used this method to gauge the plausibility of several models of rank distributions in sports performance data~\cite{morales2016generic}; for a rigorous criticism of the methods of \cite{clauset2009power} based on extreme value theory, see \cite{voitalov2019scale}.) We choose as distance metric four different test statistics~\cite{stephens1974edf}.  The first one is the standard Kolmogorov-Smirnov (KS) statistic~\cite{press1988numerical},
\begin{equation}
\label{eq:KSstat}
D = \max_{a_0 \leq a \leq a_m} | \Delta P_a |,
\end{equation}
that is, the largest magnitude of the difference  $\Delta P_a (t) = P_{\mathrm{data}}[a' \leq a] - P_a(t)$ between the cumulative distribution function (CDF) in data, $P_{\mathrm{data}}[a' \leq a]$, and the CDF of the fitted model,  $P_a(t) = \sum_{a'=a_0}^a p_{a'}(t)$, across all activities $a \in [a_0, a_m]$, where $a_0$ and $a_m$ are the minimum and maximum alter activities in the empirical ego network, respectively. The other three belong to the Cramér-von Mises family of test statistics~\cite{anderson1962distribution,choulakian1994cramer,csorgHo1996exact,lockhart2007cramer}: the Cramér-von Mises ($W^2$) statistic,
\begin{equation}
\label{eq:W2stat}
W^2 = k \sum_{a=a_0}^{a_m} \Delta P_a^2 p_a,
\end{equation}
the Watson ($U^2$) statistic,
\begin{equation}
\label{eq:U2stat}
U^2 = k \sum_{a=a_0}^{a_m} [ \Delta P_a - \langle \Delta P \rangle ]^2 p_a,
\end{equation}
and the Anderson-Darling ($A^2$) statistic,
\begin{equation}
\label{eq:A2stat}
A^2 = k \sum_{a=a_0}^{a_m} \frac{ \Delta P_a^2 p_a }{ P_a ( 1 - P_a ) },
\end{equation}
where $\Delta P_a^2$ and $\langle \Delta P \rangle = \sum_{a=a_0}^{a_m} \Delta P_a p_a$ are, respectively, the square and average of the CDF difference between model and data~\footnote{For consistency,  $\Delta P_{a_m} = 0$ and the last term in both $W^2$ and $A^2$ is set to zero.}.

\begin{table}[t]
\small
\noindent\makebox[\textwidth]{ \begin{tabular}{l r | r r r r}
\toprule
Dataset & $N$ & $n_{D}$ & $n_{W^2}$ & $n_{U^2}$ & $n_{A^2}$ \\
\midrule
Mobile (call) & 5431921 & 0.70 & 0.69 & 0.65 & 0.69 \\
Mobile (sms) & 4233187 & 0.63 & 0.69 & 0.69 & 0.63 \\
Mobile (Wu 1) & 16050 & 0.42 & 0.61 & 0.70 & 0.43 \\
Mobile (Wu 2) & 20534 & 0.38 & 0.58 & 0.68 & 0.38 \\
Mobile (Wu 3) & 4215 & 0.36 & 0.52 & 0.62 & 0.37 \\
Email (Enron) & 21984 & 0.39 & 0.49 & 0.55 & 0.40 \\
Email (Kiel) & 9842 & 0.33 & 0.42 & 0.49 & 0.34 \\
Email (Uni) & 2456 & 0.71 & 0.78 & 0.75 & 0.70 \\
Email (EU) & 866 & 0.62 & 0.58 & 0.54 & 0.60 \\
Facebook & 31429 & 0.59 & 0.65 & 0.67 & 0.59 \\
Messages & 20252 & 0.39 & 0.43 & 0.45 & 0.38 \\
Dating & 16239 & 0.56 & 0.65 & 0.67 & 0.57 \\
Forum & 4122 & 0.43 & 0.46 & 0.46 & 0.42 \\
College & 1303 & 0.65 & 0.67 & 0.66 & 0.64 \\
CNS (call) & 285 & 0.34 & 0.52 & 0.68 & 0.34 \\
CNS (sms) & 347 & 0.54 & 0.81 & 0.88 & 0.54 \\
\bottomrule
\end{tabular}}
\caption{
\small {\bf Statistical significance of maximum likelihood estimation}. Fraction $n_{\bullet}$ of ego networks satisfying the condition $p_{\bullet} > 0.1$ on the $p$-value $p_{\bullet}$ associated to the test statistics of Kolmogorov-Smirnov, Cramér-von Mises, Watson, and Anderson-Darling [$\bullet = D, W^2, U^2, A^2$, respectively; see \esref{eq:KSstat}{eq:A2stat}]. Fractions $n_{\bullet}$ are calculated relative to the number $N$ of egos in each dataset under the condition $t > a_0$ (i.e. with any level of heterogeneity on their communication signatures). The model is able to reproduce observed data for most egos, at least according to some statistic. For large datasets, statistical significance is robust to the choice of statistic.
}
\label{tab:filterStats}
\end{table}

The GOF test is as follows: Given the sample $\{ a_i \}$ from an empirical ego network, we compute the MLE $\hat{\alpha}$ numerically from \eref{eq:alphaTrascEq}, as well as the associated data statistics $D$, $W^2$, $U^2$,  and $A^2$ from \esref{eq:KSstat}{eq:A2stat}, where the model CDF $P_a(t)$ is computed numerically from $p_a(t)$ in \eref{eq:ActDistExp} with $\alpha = \hat{\alpha}$ (and $t$, $a_0$, and $a_m$ are taken from the data sample). From the model $p_a(t)$ we generate $n_{\mathrm{sim}}=2500$ simulated activity samples $\{ a_i \}_{\mathrm{sim}}$. For each simulated sample, we find its own MLE $\hat{\alpha}_{\mathrm{sim}}$ and the corresponding simulated statistics $D_{\mathrm{sim}}$, $W_{\mathrm{sim}}^2$, $U_{\mathrm{sim}}^2$,  and $A_{\mathrm{sim}}^2$. Then, the fraction of simulated statistics $\bullet_{\mathrm{sim}}$ larger than the data statistic $\bullet$ (i.e.  $D_{\mathrm{sim}} > D$ and so on) is the $p$-value $p_{\bullet}$ associated with the goodness-of-fit test, according to this particular test statistic. If the $p$-value is large enough ($p_{\bullet} > 0.1$ with 0.1 an arbitrary significance threshold), we do not rule out the hypothesis that our activity model emulates the empirical ego network. We aim at obtaining large $p$-values (rather than small),  since we want to keep the assumption that the model is a good description of the observed data (rather than reject it).

We apply the MLE fitting procedure and GOF test described above to all $N$ ego networks of each communication dataset described in \sref{sec:data} and \tref{tab:datasets} (already filtered by the condition $t > a_0$ and thus with enough data on their heterogeneous communication patterns). Then we calculate the fraction $n_{\bullet}$ of egos that satisfy the condition $p_{\bullet} > 0.1$ according to statistic $\bullet$ (\tref{tab:filterStats}), that is, the fraction of egos well described by our model of alter activity. Values of $n_{\bullet}$ vary from 33\% to 71\% for $D$ and $A^2$, and slightly increase to the range 42--88\% for $W^2$ and $U^2$.  In this sense, the ability of our model to describe empirical data is robust to the measure of statistical significance,  beyond small datasets for which both $W^2$ and $U^2$ are somewhat less restrictive.  In large datasets like Mobile (call \& sms), all four statistics imply that 63--70\% of ego communication signatures are captured by the model. Given this similarity in the behaviour of test statistics, for the remaining of results (here and in the main text) we focus on the KS statistic $D$ and drop the notation $\bullet$ in the $p$-value. 

\begin{table}[t]
\small
\noindent\makebox[\textwidth]{ \begin{tabular}{l | r r r r | r r r}
\toprule
Dataset & $N$ & $n_{\alpha}$ & $n_{\infty}$ & $n_{\emptyset}$ & $N_{\alpha}$ & $n_{CA}$ & $n_{RN}$ \\
\midrule
Mobile (call) & 5431921 & 0.70 & 0.06 & 0.23 & 3817319 & 0.95 & 0.05 \\
Mobile (sms) & 4233187 & 0.63 & 0.11 & 0.25 & 2687452 & 0.93 & 0.07 \\
Mobile (Wu 1) & 16050 & 0.42 & 0.25 & 0.33 & 6800 & 0.88 & 0.12 \\
Mobile (Wu 2) & 20534 & 0.38 & 0.27 & 0.35 & 7863 & 0.87 & 0.13 \\
Mobile (Wu 3) & 4215 & 0.36 & 0.33 & 0.31 & 1534 & 0.87 & 0.13 \\
Email (Enron) & 21984 & 0.39 & 0.34 & 0.27 & 8647 & 0.82 & 0.18 \\
Email (Kiel) & 9842 & 0.33 & 0.43 & 0.24 & 3266 & 0.83 & 0.17 \\
Email (Uni) & 2456 & 0.71 & 0.06 & 0.22 & 1746 & 0.95 & 0.05 \\
Email (EU) & 866 & 0.62 & 0.05 & 0.33 & 541 & 0.97 & 0.03 \\
Facebook & 31429 & 0.59 & 0.25 & 0.16 & 18689 & 0.83 & 0.17 \\
Messages & 20252 & 0.39 & 0.45 & 0.16 & 7814 & 0.68 & 0.32 \\
Dating & 16239 & 0.56 & 0.26 & 0.18 & 9135 & 0.85 & 0.15 \\
Forum & 4122 & 0.43 & 0.37 & 0.20 & 1762 & 0.66 & 0.34 \\
College & 1303 & 0.65 & 0.21 & 0.14 & 846 & 0.82 & 0.18 \\
CNS (call) & 285 & 0.34 & 0.29 & 0.37 & 97 & 0.85 & 0.15 \\
CNS (sms) & 347 & 0.54 & 0.07 & 0.39 & 188 & 0.99 & 0.01 \\
\bottomrule
\end{tabular}}
\caption{
\small {\bf Ego classes based on maximum likelihood estimation}.
We classify the $N$ ego networks (with $t > a_0$) in each studied dataset into a fraction $n_{\alpha} = N_{\alpha} / N$ with statistically significant MLE $\hat{\alpha}$ (relative mean activity $t_r > 0$, p-value $p > 0.1$ according to statistic $D$, and $\hat{\alpha} < \alpha_b$ with $\alpha_b = 10^3$), a fraction $n_{\infty} = N_{\infty} / N$ with infinite $\hat{\alpha}$ [\eref{eq:logDeriv} does not converge to zero below $\alpha_b$], and the remaining fraction $n_{\emptyset} = N_{\emptyset} / N$ with undefined $\hat{\alpha}$. The $N_{\alpha}$ egos with statistically significant $\hat{\alpha}$ are separated into a fraction $n_{RN} = N_{RN} / N_{\alpha}$ in the homogenous regime ($\beta < 1$), and a fraction $n_{CA} = N_{CA} / N_{\alpha}$ in the heterogeneous regime ($\beta > 1$).
}
\label{tab:filterClasses}
\end{table}

The bootstrapped calculation of $p$-values allows us to separate egos into three categories depending of the properties of the MLE $\hat{\alpha}$ (\tref{tab:filterClasses}):\\
$\bullet$ {\it Statistically significant $\hat{\alpha}$.} Out of $N$ egos with $t > a_0$ (so the empirical activity distribution is different from the initial condition of the model, see \sref{sec:MLEderiv}), in this category we only consider the $N_{\alpha}$ egos for which $p > 0.1$ (i.e. the GOF test does not rule out our model as a good description of the empirical ego network, according to the chosen statistic $D$). We also only take egos with $\hat{\alpha} < \alpha_b$, where $\alpha_b = 10^3$ is an arbitrary upper bound in the numerical calculation of MLEs via \eref{eq:alphaTrascEq}, since the derivative of the log-likelihood in \eref{eq:logDeriv} tends to 0 for $\alpha \to \infty$ (for fixed $a_0$ and $t$; see \fref{fig:figure_MLEfit}) and the numerical root search for \eref{eq:alphaTrascEq} can fail.\\
$\bullet$ {\it Infinite $\hat{\alpha}$.} From what remains, in this category we consider the $N_{\infty}$ egos for which the numerical search of the root of $F_{\alpha} - \ln ( 1 + \beta )$ does not converge below the upper bound $\alpha_b$, and we assign to them the trivial solution of \eref{eq:MLEmax}, $\hat{\alpha} \to \infty$.\\
$\bullet$ {\it Undefined $\hat{\alpha}$}. The remaining $N_{\emptyset}$ egos have an undefined $\hat{\alpha}$ and are not considered further in our analysis of alter activity.\\
\tref{tab:filterClasses} shows the relative sizes of these classes for all datasets ($n_{\alpha} = N_{\alpha} / N$, $n_{\infty} = N_{\infty} / N$, and $n_{\emptyset} = N_{\emptyset} / N$, respectively, with $n_{\alpha} + n_{\infty} + n_{\emptyset} = 1$).  Across all datasets,  33--71\% of egos (with $t > a_0$) have communication patterns well captured by the model,  while 5--45\% are compatible with the trivial fit $\hat{\alpha} \to \infty$ and in this sense belong to the homogeneous regime of uniform alter activity. Overall,  only a relatively small fraction of ego networks (14--39\%) are not well emulated by the model, a figure that improves as we increase system size.

\subsection{Activity regimes and persistence analysis in communication data}
\label{sec:activityData}

\begin{figure}[t]
\centering
\includegraphics[width=0.8\textwidth]{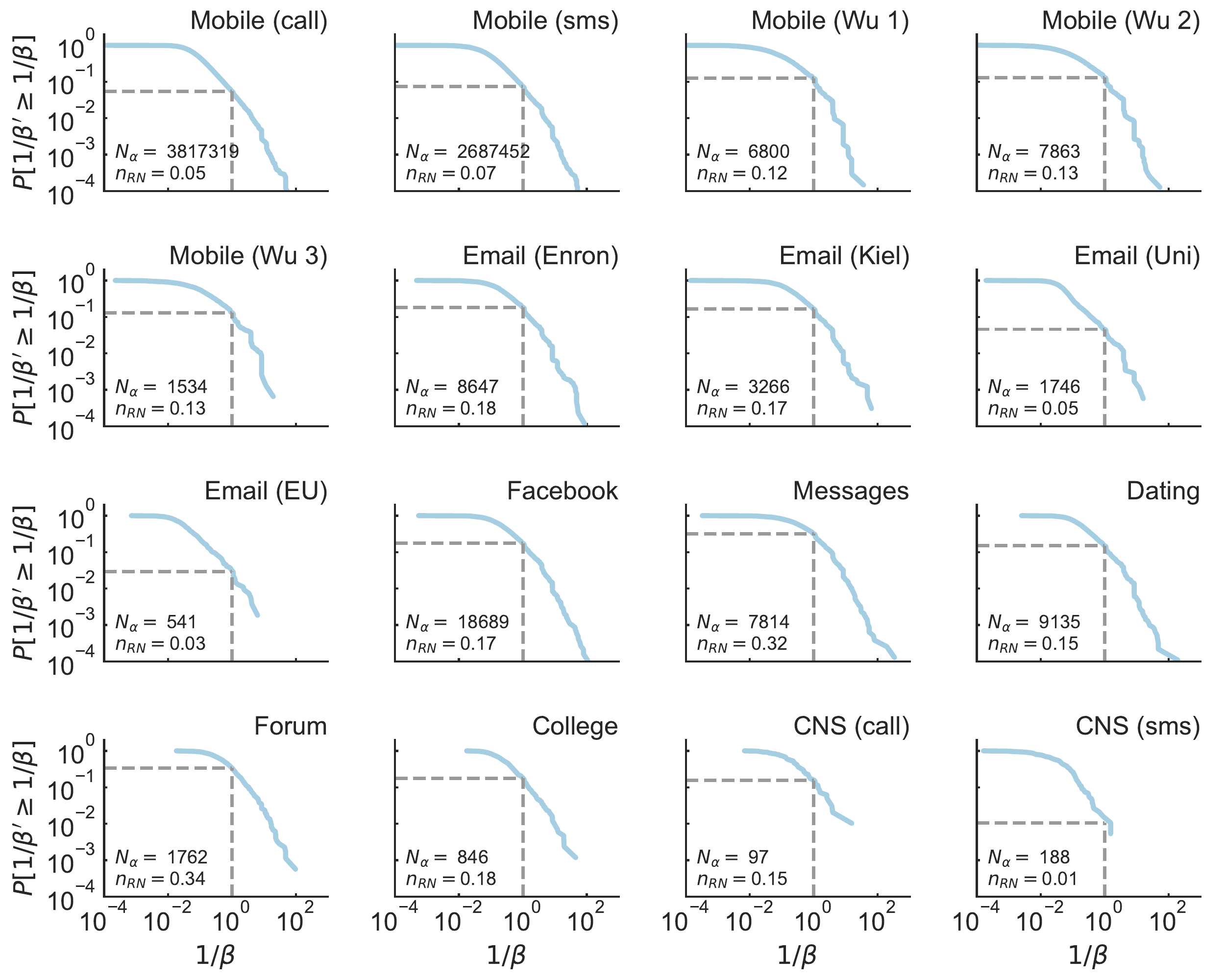}
\caption{\small {\bf Presence of cumulative advantage in communication data.} Complementary cumulative distribution function (CCDF) $P[ 1 / \beta ' \geq  1 / \beta ]$ of estimated rate $1 / \beta = \alpha_r / t_r$, fitted to the ego networks of several datasets via MLE (see \sref{sec:fitting}). After only considering egos with $t_r > 0$ (i.e. $t > a_0$), $p > 0.1$ according to statistic $D$, and $\alpha < \alpha_b$ with $\alpha_b = 10^3$, ego networks have a characteristic CCDF of the rate $1 / \beta$ with similar shape across datasets: Out of $N_{\alpha}$ egos with statistically significant $\alpha$, a relatively small fraction $n_{RN}$ distribute contacts uniformly at random among their alters ($\beta < 1$), while most egos' contact activity is more heterogeneous and centered in a few of their alters ($\beta > 1$).}
\label{fig:alphas_CCDF}
\end{figure}

\begin{figure}[t]
\centering
\includegraphics[width=0.8\textwidth]{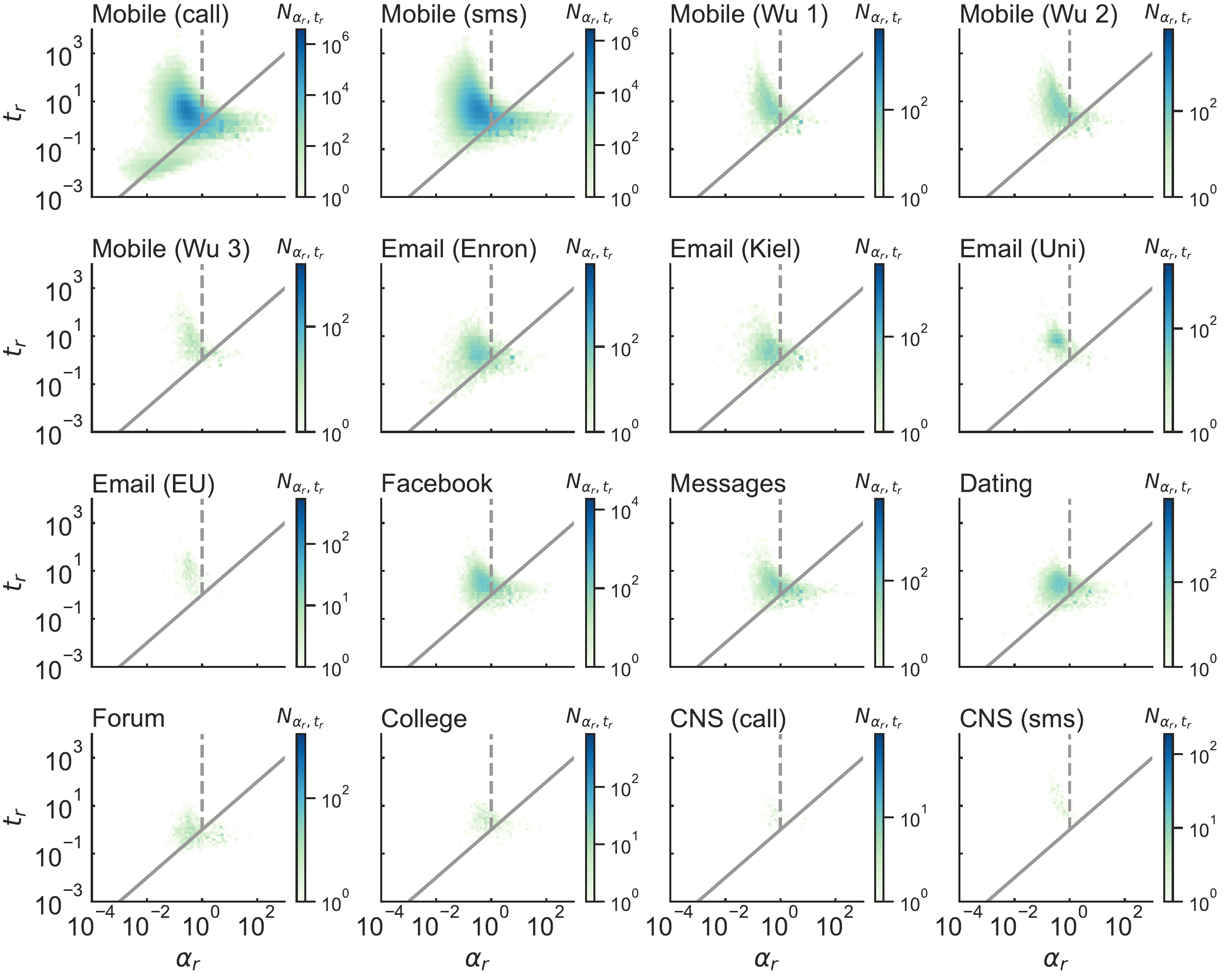}
\caption{\small {\bf Scaling phase diagram in communication data.} Hexbin histogram across datasets of number of egos $N_{ \alpha_r, t_r }$ for given values of estimated parameter $\alpha_r = \alpha + a_0$ and relative mean activity $t_r = t - a_0$, with $a_0$ the minimum alter activity. The identity $\beta = t_r / \alpha_r = 1$ (continuous line) defines a crossover between regimes of scaling of the alter activity distribution $p_a(t)$. In the homogeneous regime ($\beta < 1$; bottom), a small fraction $n_{RN}$ of egos distribute contacts at random and $p_a$ scales like a Gaussian (see \tref{tab:filterClasses}). In the heterogeneous regime ($\beta > 1$; top), $p_a$ approaches a gamma distribution and has exponential scaling. Most egos concentrate activity in a few alters ($\alpha_r < 1$; left of dashed line), while for a few egos, activity is distributed more uniformly across alters ($\alpha_r > 1$; right of dashed line).}
\label{fig:fit_corrs}
\end{figure}

The distribution of statistically significant $\alpha$ values across egos has a similar shape in all datasets, as seen from the CCDF $P[ 1 / \beta ' \geq  1 / \beta ]$ of the estimated rate $1 / \beta = \alpha_r / t_r$ (\fref{fig:alphas_CCDF}), and the associated scaling phase diagram in $(\alpha_r, t_r)$-space (\fref{fig:fit_corrs}; for comparison see \fref{fig:activity_regimes}) \footnote{For simplicity,  here and in the main text we drop the notation $\hat{\bullet}$ used for parameter estimates in \sref{sec:MLEderiv} and \sref{sec:GOFtest}.}. The way egos distribute communication events among alters lies in a spectrum: In the homogeneous regime of $\beta < 1$, a small fraction $n_{RN} = N_{RN} / N_{\alpha}$ of egos distribute activity among alters uniformly at random, with a Poissonian activity distribution asymptotically scaling like a Gaussian [see \sref{sec:randomRegime} and \eref{eq:ActDistLargeAlpha}]. In the heterogeneous regime of $\beta > 1$, a larger fraction $n_{CA} = N_{CA} / N_{\alpha}$ of egos may concentrate their contacts in a just a few of their alters, so their activity is gamma-distributed with exponential scaling for given $\alpha_r$ [see \sref{sec:gammaRegime} and \eref{eq:ActDistSmallAlpha}]. Values of $n_{CA}$ and $n_{RN}$ are shown in \tref{tab:filterClasses} and \fref{fig:alphas_CCDF}, where $n_{CA} + n_{RN} = 1$.

\begin{figure}[t]
\centering
\includegraphics[width=0.8\textwidth]{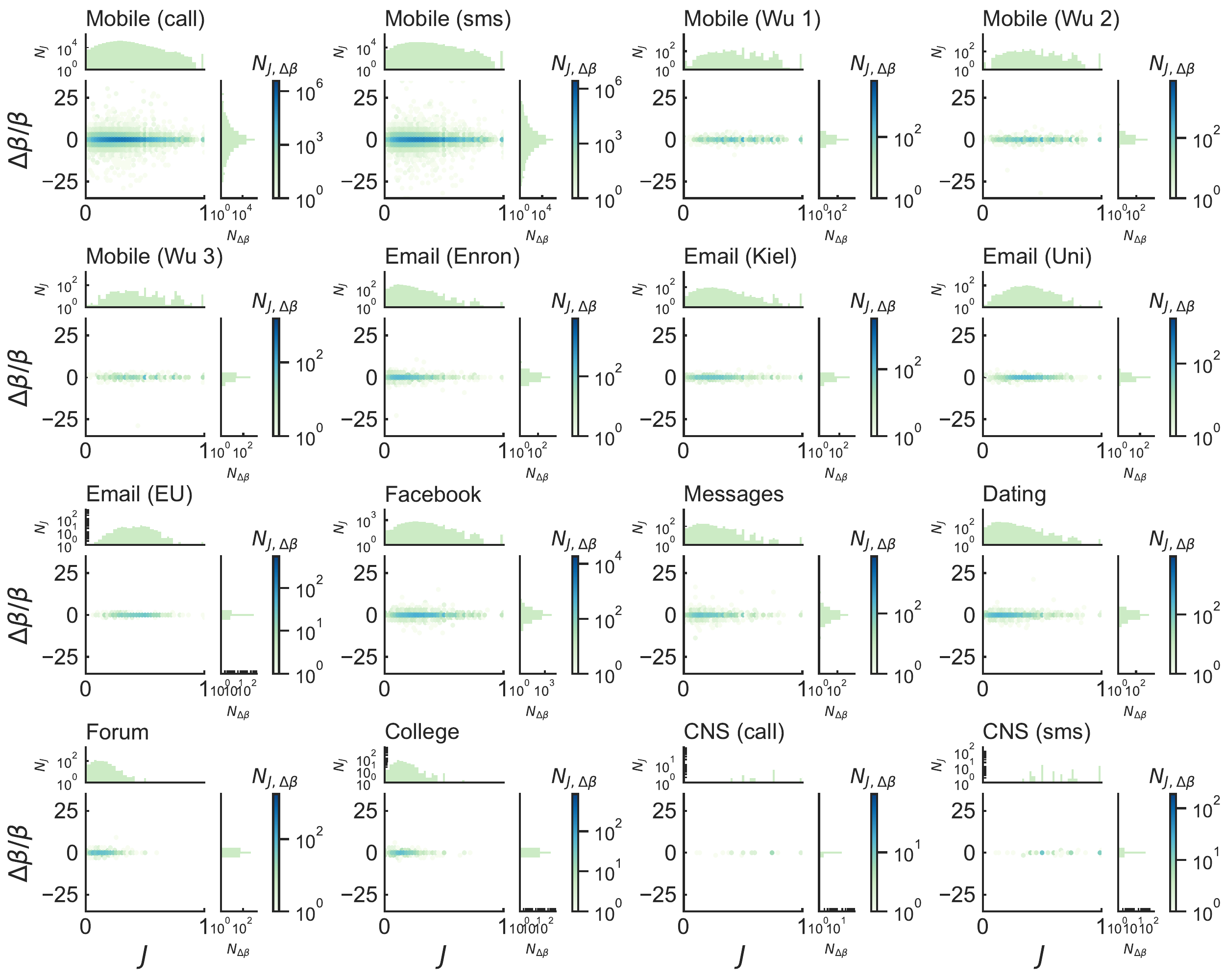}
\caption{\small {\bf Persistence of preferentiality in communication data.} Hexbin histogram across datasets of number $N_{J,  \Delta \beta}$ of egos with given alter turnover $J$ and relative preferentiality change $\Delta \beta / \beta$.  We estimate the preferentiality parameter in the whole observation period ($\beta$) as well as in two consecutive intervals of activity spanning the period ($\beta_1$ and $\beta_2$, respectively, with $\Delta \beta = \beta_1 - \beta_2$).  We also show marginal number distributions of turnover ($N_J$) and relative preferentiality change ($N_{\Delta \beta}$).  Social signatures are persistent in time at the level of individuals,  regardless of alter turnover.}
\label{fig:persistence_turnover}
\end{figure}

The presence of heterogeneous and homogeneous regimes of alter activity at the ego level, as well as the system-level similarity in the distribution of $\beta$ values across all datasets,  are both features of human communication that seem to persist in time, regardless of potential changes in the identity of alters within ego networks (\fref{fig:persistence_turnover}). To quantify this effect, we separate the observed period of activity of an ego network into two consecutive intervals with the same number of events ($I_1$ and $I_2$, see Fig. 1 in main text). We then independently estimate the preferentiality parameter for both the entire period ($\beta$) and for each of these two intervals ($\beta_1$ and $\beta_2$),  leading to a preferentiality change $\Delta \beta = \beta_1 - \beta_2$.  We also measure alter turnover as the Jaccard similarity coefficient $J = |A_1 \cap A_2 | / | A_1 \cup A_2 |$ between the sets of alters $A_1$ and $A_2$ in both intervals (with $J = 0$ implying totally different alters in $I_1$ and $I_2$, and $J=1$ exactly the same alters across intervals) \cite{saramaki2014persistence}.  \fref{fig:persistence_turnover} shows that the relative preferentiality change $\Delta \beta / \beta$ stays close to zero regardless of alter turnover,  somewhat trivially for $J \sim 1$ (since alters are anyway the same people when moving from $I_1$ to $I_2$), but remarkably also for $J \sim 0$.  In other words, the individual way in which each ego allocates communication activity among alters (driven by cumulative advantage or by random alter choice) persists in time despite potentially large changes in the identity makeup of their social networks.

{\small
\bibliographystyle{ieeetr}
\bibliography{references}
}